\documentclass[a4paper,10pt]{article}
\usepackage{ucs}
\usepackage[utf8]{inputenc}
\usepackage{amsmath}
\usepackage{amsfonts}
\usepackage{amssymb}
\usepackage{amsthm}
\usepackage{rotating}
\usepackage{xcolor}
\usepackage[english]{babel}
\usepackage[T1]{fontenc}
\usepackage{graphicx}
\usepackage{float}
\usepackage{hyperref}
\usepackage{cuted}
\usepackage{verbatim}
\usepackage{subfigure}
\usepackage{multirow}
\usepackage{bbold}
\usepackage{slashed}
\usepackage{indentfirst}
\usepackage{tcolorbox}
\usepackage{xcolor}
\usepackage{placeins}
\usepackage[a4paper, total={7in, 10in}]{geometry}

\usepackage{multicol}
\usepackage{fmtcount}
\usepackage{diagbox}

\usepackage{placeins}

\begin{document}

\title{ \bf 
Quark-antiquark states of the lightest
scalar mesons within the Nambu-Jona-Lasinio model
with flavor-dependent coupling constants }

\author{ Fabio L. Braghin  \thanks{braghin@ufg.br}
\\
{\normalsize Instituto de F\'\i sica, Federal University of Goias,}
\\
{\normalsize Av. Esperan\c ca, s/n,
 74690-900, Goi\^ania, GO, Brazil}
}

\maketitle

\begin{abstract}
The quark antiquark components of the  U(3)
lightest  scalar meson   nonet 
 are investigated by considering 
the Nambu-Jona-Lasinio  model with flavor-dependent coupling constants
 that were derived by
considering (non-perturbative) gluon  exchange.
The strange quark effective mass ($M_s^*$)
 is varied such that the
strangeness content    of these scalar mesons can be 
analyzed further.
  The neutral   states  $S_0$, $S_8$ and $S_3$, are adopted as the most
relevant quark-antiquark components respectively of the 
$\sigma(500)$, $f_0(980)$ and $A_0^0(980)$ mesons.  
As a result, the  mass hierarchy between 
states $S_0$ and $S_8$,
and  $K^*_0(700) - A_0(980)$ mesons, can be
respectively  inverted and corrected for
quite low values of the  strange quark constituent mass.
Besides that, for some  particular values of  $M_s^*$,
 the  masses of $\sigma(500)$ and $f_0(980)$
can also be obtained
by considering 
the mixing coupling constant $G_{08}$.
However, the masses of all the nine mesons 
are not  described simultaneously  in a self-consistent procedure and 
this goes along with 
the need of non-quark-antiquark states to  completely  describe their masses.
A neutral-meson mixing matrix  is defined
and the leading mixing angle is found by 
fitting a
 correct value of 
 the $\sigma (500) -f_0 (980)$ mass difference.
Two different estimates  for the
strengths of the mixings $A_0^0 \to f_0$ and $f_0  \to A_0^0$ 
are proposed,  leading to somewhat different behaviors.
Firstly by assuming leading  transitions to intermediary 
flavor eigenstates and secondly by considering a dynamical 
evolution. 
Results may be 
 in good agreement with 
  experimental values  from BESS-III  depending on the value of the 
strange quark effective mass.
\end{abstract}

\section{ Introduction }

The  structure
 of the 
 lightest scalar mesons - 
$\sigma(500)$ (or $f_0(500)$), 
 isodoublets $K_0^*(700)$ (or $\kappa(700)$),
 isovector $A_0^{0,\pm}(980)$ and 
$f_0(980)$ \cite{PDG}  - 
is among the major challenges for the quark model.
The   usual  description of the lightest scalar mesons as   quark-antiquark states of 
a flavor U(3) nonet presents several difficulties
and it  is not believed to be correct  by many groups, in spite of being
 a highly controversial subject  \cite{pelaez-review}.
However, these mesons possibly form a nonet \cite{maiani-etal04,close-tornq,eLSM}
and this  is seemingly required or supported by fundamental properties 
 \cite{pelaez-review,oller-mixing}.
Different structures have been envisaged   to describe these  mesons,
  such as   tetraquarks, mesons molecules or 
mixed states that may include also  glueballs.
A  list of works addressing the subject  is very long, 
for a few  examples see Refs.
\cite{PDG,maiani-etal04,scadron-etal,jaffe,pelaez-status,su-etal-npa2007,brigitte-etal,giacosa-etal2018,dmitrasinovic,scalars-1,scalars-2,scalars-2b,scalars-3,WGR-2015,scalars-molecular,close-tornq}.
For a broader review on the subject, the reader can see  \cite{pelaez-review}.
Further exotic structures contributing, including 
dark matter, are found for example in  Ref.
 \cite{lattic-for-DM}.
Further experimental investigation is done and planned, for example
in  LHCb, BESS-III, COMPASS, GLuex, FAIR-GSI  
\cite{BESS-III-mesons,BESIII,fair,compass,gluex}.
Mixing between different types of states is believed to play an important role 
although there might have a difference between mixing of 
bound states and  of unbound states.
This work considers mostly  that all these   scalar quark-antiquark
states
are  bound, and  mixings may occur
between  bound states.

 The structure of the  mesons  $\sigma(500)$ and $f_0(980)$ 
are  the most controversial ones.
From the point of view of an usual constituent quark model, 
their masses nearly cope respectively  with
 $M_\sigma \sim 2 M_u^* \sim (M_u^*+M_d^*)$,
or slightly smaller,  and
$M_{f_0} \sim 2 M_s^*$. 
Several works, usually based on  descriptions with the Linear Sigma Model 
 (LSM) or 
quark-level Linear Sigma Model (QLLSM) 
suggest that the lightest scalars
can be indeed accommodated in a flavor nonet 
\cite{close-tornq,schumacher,scadron-etal,qqlsm1,qqlsm2}.
This picture however is not corroborated by several investigations 
  which present several  indications
they should have different structure, maybe more important
\cite{pelaez-review}.
In particular, the "discovery" of the  $\sigma$ meson 
was  very troubled  in the early investigations of the $\pi-\pi$ scattering
 mostly because of its very large decay width,
being somewhat above the $\pi-\pi$ scalar channel threshold 
and considerably below the kaon-kaon threshold.
During years, it was believed 
that  $K_0^*(700)$ would be different from the  $\kappa$,
 whose existence has also  been challenged \cite{close-tornq}.
Nowadays, they are seen  to be the same particle
\cite{PDG}.
The $f_0(980)$ lies very close to the $K-\bar{K}$ threshold, making more difficult 
to isolate or to test both  quark-antiquark and molecular structures at once.
By analyzing the $f_0(980)$ production at LHCb, in color evaporation and color-singlet gluon-gluon fusion,
it was  suggested that it is not a color singlet
\cite{prod-f0-lhc-nosinglet}.
Besides that,
 $f_0(980)$ is nearly degenerate with the isotriplet $A_0(980)$.
Whereas the sigma-meson  is  usually suggested 
to not have $\bar{s}s$ strangeness,
as found from Breit-Wigner parameterization
 \cite{strange-f0}, 
the $f_0(980)$ seems  to have an important strangeness content
that often may be incorporated, supplemented  or represented in a kaon molecule 
or even considered to be a  $\bar{s}s$ state or  an octet state 
\cite{pelaez-review,1808.08843}.
Within the   large Nc limit  and/or with Regge theory 
the sigma should   not be only a $\bar{q}q$ state 
\cite{pelaez-review,largeNc-scalars,largeNc-scalars2}
and, 
currently, scalars $\sigma$ and $f_0$ are mostly interpreted as possessing 
tetraquarks as main components \cite{jaffe,lhcb,pelaez-review}.
 Both $\sigma$ and $f_0$ can be generated dynamically, for example,
from the coupled channel scattering amplitudes in the chiral 
unitary framework \cite{sigma-f0}.
Scalars $f_0(980)$, and  also $A_0(980)$, were found to be compatible 
respectively with 
 larger molecular 
and  quarkonia components \cite{wang-etal}.
However, in general, the isotriplet $A_0^0(980)$ seems rather a flavor  counterpart of the pseudoscalar pions.
Further support for larger components of octet-type
for  $f_0(980)$ and singlet-type for  $\sigma(500)$ 
was given  in Ref. \cite{klempt} 
in which several spin-partners were proposed.
For that, a mixing angle between flavor eigenstates was 
calculated to be
$\theta_s = 14 \pm 4 ^{\circ}$
and then  mixings to tetraquarks,
meson-meson and glueball states were envisaged.
Besides this fragrant  $\sigma-f_0$ problem,
several  theoretical estimations for the
 masses of 
 $A_0(980) ( I(J^C) = 1(0^+) )$ and $K^*(700) ( \frac{1}{2}(0^+) )$
  were  found to provide a wrong
 mass hierarchy 
 \cite{pelaez-status,pelaez-review}.

All these  problems may manifest 
 in the {\bf   standard version}  of the Nambu-Jona-Lasinio (NJL) model 
\cite{NJL,klevansky,vogl-weise,klimt-etal-A,schechter-etal,hiller-etal2007,osipov-etal-1,osipov-etal-2,volkov}.
The NJL model is an emblematic low energy QCD model
 that reproduces many
low energy (global) hadron  observables along the (constituent)  Quark Model
 by incorporating 
 chiral symmetry properties and  theorems  of QCD 
\cite{cheng-li,IZ,schwartz}.
It usually  provides an interesting framework 
to  extend the (constituent)  quark model.
Dynamical chiral symmetry breaking (DChSB) is one of its important features,
according to which chiral scalar quark-antiquark condensates contribute largely to the 
hadron masses and properties.
 As a result, pions (and kaons) emerge as (quasi) Goldstone bosons
in agreement with phenomenology which shows an overwhelming role for these particles,
 whose 
masses are very small when compared to typical hadron masses.
Besides that, the NJL model is a  non-renormalizable model 
and its   calculated observables
depend on a chosen ultraviolet (UV) cutoff. 
Although
 results from the NJL model
are known to depend on the chosen regularization scheme,
 it has been found in different works 
that the difference among the different schemes for many observables
of the light quark sector
  is quite small
\cite{klevansky,kohyama-etal}.
The determinantal, instanton induced,
 't Hooft type interaction, 
  appears due to the axial anomaly $U_A(1)$, and it is often incorporated 
\cite{UA1}.
For  flavor $N_f\times N_f$,
it  can be written as:
 ${\cal L}_{tH} = \kappa \left( \det (\bar{\psi} P_L \psi ) 
+ \det (\bar{\psi} P_R \psi ) \right)$,
  where $P_{R/L} = (1 \pm \gamma_5)/2$ are the chirality projectors and $\kappa$ is a coupling
constant  usually taken as a free parameter of the model.
This interaction has an important role for the description of the 
$\eta-\eta'$ puzzle by means of mixings,
 although it has been found that 
flavor dependent coupling constants also contribute for that \cite{PRD-2021,JPG-2022}.
This 't Hooft determinantal interaction, and sometimes
higher order interactions,  can help to describe the light scalars in a U(3) nonet
\cite{su-etal-npa2007,osipov-etal-1,osipov-etal-2,hiller-etal2007,scalars-cheng-etal,kuroda-etal,klevansky,vogl-weise,hatsuda-etal}.
Some versions of this extended  model, however, 
do not reproduce the lightest scalar mesons as quark-antiquark states
but slightly heavier scalars with masses around 1.0-1.5GeV
\cite{dmitrasinovic}.
In the present work, although we assume that axial anomaly 
takes place, this determinantal interaction will not be considered so that
the effect of  flavor-dependent coupling constant may be
isolated and emphasized.
Furthermore, the concern in the present work is not  really to  
present a model that reproduces the masses of
 the lightest scalars but to investigate how the scalar sector of the model  behaves
with respect to the flavor-dependent coupling constants
and with  the values of the strange quark mass.
 One important consequence 
of the polarization-induced coupling constants
is that 
the quark effective masses undergo further modifications, in particular 
the strange quark effective mass $M_s^*$ is lowered considerably.

Mixings are believed to be an important effect in the
scalar and pseudoscalar   mesons structure.
Mixing matrices for $\sigma(500)$ and $f_0(980)$
 with glueball or as tetraquarks  were proposed, for example, in
\cite{klempt,f0-glueball,mix-qcdsr}.
However, it is usually argued that 
the lightest scalar $0^{++}$ glueball
 seemingly has a too large mass to make possible a 
non-negligible mixing with light scalars
 \cite{f0-glueball,close-tornq}.
  BES III Collaboration  observed, for  the first time, clear signs of meson mixing
between $A_0^0(980) -f_0(980)$, for which 
 different intensities for 
$f_0\to A_0$ and $A_0\to f_0$ 
  were identified
\cite{a0-f0-mix-exp}.
In the present work, only mixing between the quark-antiquark states will be addressed.

In this work, the  flavor-dependent NJL-model  as
presented in Refs. \cite{PRD-2021,JPG-2022}
is considered for the investigation of the quark-antiquark components of the 
lightest scalar mesons arranged in a U(3) nonet.
The   fit of parameters, with 
corresponding self-consistent calculation, will be taken from 
Ref. \cite{JPG-2022} where  the values of the 
 neutral  pion and kaon masses were used 
as fitting parameters and several observables were  calculated.
Whereas in Ref. \cite{PRD-2021}
a perturbative investigation of the effect 
of the flavor-dependent coupling constants $G_{ij}$ was done,
in Ref. \cite{JPG-2022} a slightly different approach was 
used by considering self-consistent values of quark effective masses 
and $G_{ij}$.
Results for the light pseudoscalar mesons masses
 and properties calculated  in these two  ways  are very similar.
These flavor dependent coupling constants
  are determined by the 
quark effective masses
and   they  have been calculated 
by considering the effect of gluon exchange, being therefore 
dependent on an effective gluon propagator.
At the end of the calculation, one might have 
stronger and somewhat amplified  flavor dependent effects dictated solely 
by the current quark masses that are the only QCD parameters that 
control flavor asymmetry \cite{gasser-leutwyler,donoghue}.
The first  effect of the flavor dependent coupling constants is to provide lower
strange quark effective  mass, being still 
   larger than the up and down ones.
Besides that,  two types of mixings appear in this description:
a quantum mixing, due to the different representations 
of the flavor group
needed to 
describe quarks and mesons, and mixing interactions due 
to the polarization tensor defining mesons structure that leads to $G_{i\neq j}$
($i,j=0,...N_f^2-1$) and 
$G_{f\neq g}$ ($f,g=u,d,s$).
The mixing type interactions $G_{i \neq j}$, $G_{f_1\neq f_2}$
 are 
proportional to quark effective mass differences
and therefore they  have considerably smaller numerical values.
 The only  mixing interactions $G_{i \neq j}$  considered will be those needed to 
calculate explicitly the mixing between states $S_0-S_8$ and $S_3$.
They  
contribute to the  $\sigma(500)-f_0(980)$ mass difference 
and to the mixing $A_0(980)-f_0(980)$.
The possible mixing $A_0(980)-\sigma(500)$, emerging from the 
coupling constant $G_{30}$ for the states $S_3, S_0$
 is the smallest one, and it  will not be calculated.
Besides that, the strange quark effective mass $M_s^*$ will be varied
 in two ways, as proposed in Ref. \cite{JPG-2022}.
Firstly, this variation will be done  self consistently (s.c.), i.e. 
by varying the  strange quark current mass and by 
calculating all the effective masses from the gap equations.
Secondly, $M_s^*$ will be varied freely 
by keeping the up and down effective masses calculated self consistently.
The following quark-antiquark states will be adopted for the 
  components of the lightest mesons
 \cite{maiani-etal04,hiller-etal2007}:
\begin{eqnarray} \label{scalars-structure}
&& A_0^0 \sim \frac{1}{\sqrt{2}} (\bar{u} u - \bar{d} d)
, 
\;\;\;\;\;\;
A_0^\pm \sim \bar{u} d , \bar{d} u
,
\;\;\;\;\;
K^*_0, \bar{K}^*_0 \sim  \bar{d} s , \bar{s} d
,
\;\;\;\;\;
K^{*\pm}_0 \sim \bar{s} u , \bar{u} s
,
\\
&&
\sigma \sim   \frac{1}{\sqrt{3}} (\bar{u} u + \bar{d} d + \bar{s} s)  
, 
\;\;\;\;\;\;\;\;\;\;\;\;\;\;
f_0 \sim \frac{1}{\sqrt{6}} (\bar{u} u + \bar{d} d - 2 \bar{s} s).
\end{eqnarray}
The work is organized as follows.
In the next section, the NJL model with flavor dependent coupling constants is presented
and the so-called bosonization procedure  is applied by means of the auxiliary field method for 
quark-antiquark states, $S_i$ with $i=0,1...8$.
This usual procedure  leads to gap equations for the 
effective masses $M_f^*$  and bound state equations, i.e. Bethe-Salpeter  at the Born level (BSE), for the flavor eigenstates $S_i$.
Most of the  solutions presented will be uncoupled, i.e. without mesons mixings.
A coupled solution for the BSE of the  $S_0,S_8$  
states will be investigated since this
is the leading mixing interaction ($G_{08}$), being the others ($G_{03},G_{38}$)
considerably smaller.
The role of mixing interactions between the flavor eigenstates $S_0-S_8$
 will be worked out 
 further, inspired by the corresponding treatment for the lightest pseudoscalar mesons.
The leading mixing $S_0-S_8$  will be assumed to  describe
the mass difference of scalars   $\sigma(500)$ and $f_0(980)$ mesons.
By assuming the leading flavor-eigenstate content of the neutral  mesons 
$A_0(980),\sigma(500),f_0(980)$,
an estimation for the strength of the mixings 
$f_0 \to S_3 \to  A_0^0$ and $A_0^0 \to S_8 \to f_0$ will be provided.
 A second estimation of these $f_0-A_0$ mixings will be provided by means of a 
standard dynamical way.
In section (\ref{sec:numerics}) numerical results will be present 
for a three-dimensional UV cutoff and in the last section 
a summary with conclusions.

\section{ Flavor dependent NJL and quantum mixing}

The  Lagrangian of the NJL model with flavor dependent coupling constants 
 can be written as:
\begin{eqnarray} \label{Lagrangian}
{\cal L} &=&  \bar{\psi} R_0^{-1} \psi + 
\frac{G_{ij} }{2}
[ ( \bar{\psi} \lambda_i \psi ) ( \bar{\psi} \lambda_j \psi )
+  
 ( \bar{\psi} i\gamma_5 \lambda_i \psi ) 
( \bar{\psi} i \gamma_5\lambda_j \psi )
] ,
\end{eqnarray}
where 
$R_0^{-1}=( i \slashed{\partial} - m_f )$, and the flavor indices are 
 $_{f  =  u,   d,   s}$ and  
 $i,j=0,...N_f^2-1$, 
being $N_f=3$ the number of flavors, and $\lambda_i$ 
are the flavor Gell-Mann matrices
with  $\lambda_0 = \sqrt{2/3} I$.
The mass matrix $m_f$ is diagonal with the up, down and strange current masses
usually fixed by values close to the ones in PDG \cite{PDG} such that values of observables, 
as for example light mesons masses, are reproduced.
In this version of the flavor dependent model, the  chiral symmetry breaking
induced by the one-loop derivation of $G_{ij}$, and that leads
to different scalar and pseudoscalar coupling constants,
was neglected, i.e. it will be considered that
 $G^s_{ij} = G^{ps}_{ij}$.
As noted in Refs. \cite{PRD-2021,JPG-2022}
the scalar coupling constants have a quite characteristic behavior with the 
flavor asymmetry (u-d and u-s) that produces considerable deviations 
with respect to the expected values of the pseudoscalar mesons observables.
It is interesting to note that for strong magnetic fields
the opposite behavior  
was found, being the scalar $G_{ij}$  more
suitable. 
In that case, the 
pseudoscalar coupling constants produce the wrong behavior of 
the neutral pion and kaon masses  as functions of the magnetic field
\cite{PRD-2022b}.
The following important properties, 
due to CP and electromagnetic U(1) invariances, hold:
$G_{22} = G_{11}$, 
$G_{55} = G_{44}$,
$G_{77} = G_{66}$
and also $G_{i j} = G_{j i}$.

These coupling constants are calculated for a given NJL coupling constant
of reference - $G_0 = 10$ GeV$^{-2}$ - at the one loop level. 
This value of the NJL coupling constant is slightly larger than 
values usually adopted for the NJL model that provide results  consistent with 
light mesons phenomenology.
These values are however smaller than values
consistent with an identification of the NJL model with the QLLSM \cite{schumacher}.
As explained in \cite{JPG-2022} this value provides faster convergence for the
self-consistent calculation of 
the quark effective masses and coupling constants $G_{ij}$.
However, instead of doing the strict one-loop calculation with the NJL interaction,
 a
quark-antiquark interaction mediated by a non perturbative one gluon exchange
 had been considered.
Effective gluon propagators (EGP)  with different normalization were considered
and therefore a normalization is needed such that 
the effect of flavor-dependent corrections
can be as properly evaluated and compared  with respect to the standard NJL - with $G_0$.
These diagonal coupling constants $G_{ii}$  have been normalized
such that, for varying quark (effective) masses, they 
all reproduce the value of reference, $G_0 = 10$ GeV$^{-2}$, when
$M_u^*=M_d^*=M_s^*$ \cite{JPG-2022}.
Although the adopted notation does not indicate it explicitly,
all the coupling constants used in this work are the normalized ones,
i.e. $G_{ij}^n$.
There emerges a self-consistent set of equations - gap equations and 
coupling constants equations - that were solved numerically.
For that, the starting point is the standard NJL model with $G_0$.
The corresponding gap equations - obtained by neglecting all
explicit mixing interactions $G_{i\neq j}$ and $G_{f_1 \neq f_2}$
-  were considered to be given  by:
\begin{eqnarray}  
 \label{gap-g2}
 {M_f^*} - m_f = G_{ff} \; Tr \; (S_{0,f}) \; + \; {\cal O} (G_{mix}),
 \end{eqnarray}
where
 $Tr$ is the generalized trace in flavor, color, Dirac and momenta.
 $G_{ff}$ are  written below, the quite smaller $G_{mix}$ are neglected 
and  $S_{0,f} = (i \slashed{\partial} - M^*_f)^{-1}$ stands for the quark propagator 
in the presence of the DChSB with 
the effective mass $M^*_f  = m_f  + \bar{S}_f$, being   $\bar{S}_f$
the average value of the scalar auxiliary field in the vacuum.
By making the coupling constant to be the 
original NJL coupling constant, $G_{ii} \to G_0$, these gap equations
reduce  to the standard ones with $G_{ff} =  G_0$.
However, the calculation of the $G_{ij}$ depends on the effective masses
that are solved from $G_{ff}$ being that  $G_{ij}$ and $G_{ff}$ 
act in different representations.
Therefore, a quantum mixing effect arises and this may yield 
an amplified contribution of the strange sea quarks in to the dynamics of the 
up and down quarks.
Although  pseudoscalar (meson)  states will not be considered in the
present work, the parameters of the model used in the numerical
calculations were fixed in Ref. \cite{JPG-2022}
by performing
a self-consistent calculation was done for the effective masses
$M_f^*$  and coupling constants $G_{ij}$.

 The change of  representation, for $G_{i\neq j}=0$, for the 
neutral diagonal couplings $i=0,3,8$, brings the following relations:
\begin{eqnarray} \label{G-K}
 2 G_{uu}
  &=&
 2  \frac{ G_{00}^n }{3}
 + G_{33}^n + \frac{G_{88}^n}{3} 
,
\nonumber
\\
2 G_{dd}
 &=& 2 \frac{ G_{00}^n}{3} 
 + G_{33}^n + \frac{G_{88}^n}{3} 
,
\nonumber
\\
2 G_{ss} 
  &=& 2 \frac{ G_{00}^n }{3}
+ 4  \frac{G_{88}^n}{3}  .
\end{eqnarray}
In Refs. \cite{PRD-2021,JPG-2022} it was introduced 
a small parameter $x_s$ that controls
the strength of the interaction $G_{88}$ in these relations, 
such that the strength of the quantum mixing represented in these
relations can be measured in two different ways.
$G_{88}$ provides the main contribution of the strange quark to the 
other lighter channels.
 Although it induced a small difference in the results,
the difference was reasonably small and these parameters will be    $x_s=1$.
 The  mixing  coupling constants $G_{i\neq j}$  ($i,j = 0,3, 8$) will be seen 
to produce mixing between flavor eigenstates $S_0, S_3, S_8$
which manifest in neutral mesons mixings $\sigma(500)-f_0(980)-A_0(980)$
as defined above.
For instance, the larger one is 
$G_{08}$ that  mixes  states  $S_0$ and $S_8$
and this should contribute to the description of 
two  mesons,
 $\sigma$ and $f_0$.

 \subsection{Quark-antiquark  states of scalar mesons}
\label{sec:BSE}

The auxiliary field method (AFM) is  suitable to
describe flavor-multiplets of the usual  quark model 
 for quark-antiquark states 
by 
associating each  scalar quark current to a given 
scalar quark-antiquark state $S_i$.
This is naturally implemented 
by means of  the following standard unit integral 
  in the generating functional the scalar  
auxiliary fields are  introduced by neglecting all the explicit mixing interactions:
\begin{eqnarray}
1 = \int D [S_i]  exp^{ - \frac{i}{2 {G}_{ii} }
\int 
(S_{i} - {G}_{ii} j_{i}^s )
 ( S_{i} - {G}_{ii} j_{i}^s ) } .
\end{eqnarray}
The same procedure is usually  considered for the pseudoscalar mesons 
arranged in a U(3) nonet.
 The following
 parameterization of the quark-antiquark components of  the 
light scalar nonet will be considered:
\[ 
 S_a \frac{\lambda_a }{\sqrt{2}}   
=
 \left( \begin{array}{c c c }
\frac{ a_0^0 + \frac{ \sqrt{2} S_0 + S_8  }{\sqrt{3}} }{\sqrt{2}} & a_0^+ & \kappa^+
\\
a_0^-  & \frac{- a_0^0 + \frac{ \sqrt{2} S_0 + S_8  }{\sqrt{3}} }{\sqrt{2}} & \kappa^0
\\
\kappa^-  & \bar{\kappa}_0 & \frac{\sqrt{2} S_0 - 2  S_8}{\sqrt{6}}
   \end{array} \right)
.
\]
\begin{eqnarray} \label{S0S8}
\mbox{ where} && 
 |S_0 > =   \frac{1}{\sqrt{3}}
| \bar{u}u + \bar{d} d + \bar{s} s >
,
\;\;\;\;
|S_8 > =  \frac{1}{\sqrt{6}}
| \bar{u}u + \bar{d} d - 2 \bar{s} s >
\end{eqnarray}

The resulting model for scalar
auxiliary  fields can be written as
\begin{eqnarray}   \label{exp-1}
S_{d} &=& 
- \int_x  \frac{1 }{2 G_{ii} } \left[  S_i (x) S_i (x) 
\right]
- i 
Tr \ln
\left[
1 + 
{S}_{0,f} 
\lambda_i S^i  (x) 
\right]  + C_0,
\end{eqnarray}
where
$\int_x = \int d^4 x$, $C_0$ is an irrelevant constant
and $Tr$ 
stands for the generalized trace in all internal indices and spacetime.
From this effective action the gap equations 
are obtained as 
saddle point equations.

The bound state equation (BSE) for the present NJL model
is a  Bethe-Salpeter equation at the Born level.
For  the case of flavor dependent coupling constants,
these BSE are  given by:
\begin{eqnarray} \label{BSE-1}
0 &=& 1 - 2 {G}_{ij} \Pi_{f_1f_2}^{ij} (P_0^2=M_{S}^2, \vec{P}^2) ,
\\
\label{polariza-tensor}
\Pi_{f_1f_2}^{ij} (P_0, \vec{P}=0)
&=&
i Tr_{D,F,C}
\int \frac{d^4 k}{(2\pi)^4} 
\lambda_i
S_{0,f_1} (k+ P/2) \lambda_j S_{0, f_2}(k- P/2),
\end{eqnarray}
where $tr_{F,C,D}$ stands for the traces in flavor, color and Dirac 
indices.
Note that the indices $i,j$ of
the Gell-Mann matrices - adjoint representation -
 are tied with the indices $f_1,f_2$ of the 
fundamental representation of the quark propagators
for each particular channel in the  integral and in the coupling constants $G_{ii}$.
In the scalar  channel,  the following usual association will be considered
for the isotriplet and iso-doublet states  mesons:
\begin{eqnarray}  \label{A0K0}
A_0(980): \;\;  i=1,2, \;\; \mbox{with} \;  f_1,f_2=\bar{u}d / \bar{d}u;
&& 
\mbox{and} \; i=3 \; \mbox{ with} \;\; f_1,f_2= \bar{u}u+\bar{d}d,
\\
K_0^*: \;\; 
i=4,5 \;\;  \mbox{ and}  \;\; f_1,f_2=\bar{u}s / \bar{s}u,
&& 
\mbox{and} \;\; i=6,7 \;\; \mbox{with}
\;\; f_1,f_2= \bar{d}s/\bar{s}d.
\end{eqnarray} 
All types of mixings
will be  also neglected in the BSE, except in one situation described below.
Therefore, the three polarization tensors for the neutral (diagonal) 
states, by neglecting mixing terms, will reduce to:
\begin{eqnarray} \label{modes038}
\Pi_{f_1f_2}^{00}  = 
\Pi_{00}  &=& i \frac{8 N_c}{3} ( \Pi_{uu} +  \Pi_{dd} + \Pi_{ss} ),
\\
\Pi_{f_1f_2}^{33}
= \Pi_{33}  &=&  i 4 N_c ( \Pi_{uu} +  \Pi_{dd}),
\\
\Pi_{f_1f_2}^{88}
= \Pi_{88}  &=& i  \frac{4 N_c}{3} ( \Pi_{uu} +  \Pi_{dd} + 4 \Pi_{ss} ),
\end{eqnarray}
where $\Pi_{ff}$ are  the resulting quark-antiquark components from
Eq. (\ref{polariza-tensor}).
The modes $S_0$ and $S_8$ usually may  be  expected to be associated 
to the quark-antiquark components of the
 mass eigenstates $\sigma (500), f_0(980)$
and  a mixing interaction $G_{08}$ will be considered for  that.

The above BSE's for the  scalar quark-antiquark states 
have   quadratic UV   divergent integrals that are the same present in the 
gap equations.
By eliminating these quadratic divergent integrals in favor of the 
solution of the gap equations,  the resulting BSE can be written  as:
\begin{eqnarray} \label{BSE-Gff-ii} 
 (P^2 -  ({M_{f_1}^*}  + {M_{f_2}^*}) ^2 ) 
{G}_{ij} I_{2,f_1f_2}^{ij} &=&
\frac{{G}_{ij}}{2 }  \left( 
\frac{1}{ \bar{G}_{f_1f_1}} \frac{m_{f_1}}{M_{f_1}^* }
+
\frac{1}{ \bar{G}_{f_2f_2}} 
 \frac{m_{f_2}}{M_{f_2}^* }
\right)
+ 1 - \frac{{G}_{ij}}{2} \left( 
 \frac{1}{\bar{G}_{f_1f_1} } 
+ 
 \frac{1}{\bar{G}_{f_2f_2} }
\right)
,
\end{eqnarray}
where the integrals 
$I_{2,f_1,f_2}^{ij}$ are the log divergent parts of the polarization tensors 
$\Pi_{f_1 f_2}^{ij}$,
when $ij \neq 00,33,88$, being given by:
\begin{eqnarray}
I_{2,f_1,f_2}^{ij} = 4 N_c \int \frac{d^4 k}{(2 \pi)^4}
\frac{1}{ (k+Q/2)^2 - {M_{f_1}^*}^2 } \frac{1}{(k-Q/2)^2 - {M_{f_2}^*}^2}.
\end{eqnarray}
These integrals have additional singularities for external energy larger than
the sum of the two quark effective masses, i.e. $P_0 > (M^*_{f_1} + M^*_{f_2})$
leading to the (usually non-physical) decay of the meson bound state.
An infrared-cutoff, $\Lambda_{IR} \sim 120$MeV, can be used 
such that this problem is avoided \cite{cutoff-IR}.
In these equations (\ref{BSE-Gff-ii}), the coupling constants  $\bar{G}_{ff}$  
come from the gap equations. 
It turns out however that the coupling constants in the 
 BSE Eqs. have a different  
renormalization than the ones in the GAP equations 
- as shown explicitly in Ref. \cite{IJMPA-2022}   for a different model.
Therefore, it was  suggested in Ref. \cite{JPG-2022},
that the flavor dependent coupling constants in the BSE $G_{ij}$
 may assume different
normalization values from  the gap equations,
i.e. ${G}_{ff}$.
The following choice done was $\bar{G}_{ff} = G_0$.
and
results for pseudoscalar mesons observables, 
by performing  the self-consistent calculation for $G_{ij}-M^*_f$ \cite{JPG-2022},
become very similar to the perturbative calculation 
of Ref. \cite{PRD-2021}.
Preliminary results for  considering $\bar{G}_{ff}= G_{ff}$  
indicates the fitting procedure has to be done again
and the  values of the model parameters, quark current masses and cutoff,
would change somehow by maintaining the same behavior of results/observables.
This will not be exploited in the present work.

Next, the equations for the masses of the quark-antiquark 
  components of  
 $\sigma(500)$ and $f_0(980)$ mesons will be presented for 
the coupled  modes  $S_0$ and $S_8$.
By neglecting the (considerably smaller) mixings to the state $S_3$,
 the solutions
for Eqs. (\ref{BSE-1}) 
can be obtained by solving the following algebraic equations:
\begin{eqnarray} \label{BSE-mix08}
0 &=&
1 - \left( G_{00} \Pi_{00} + G_{88} \Pi_{88} + 2 G_{08} \Pi_{08} \right)
\pm 
\sqrt{ D_{08} } ,
\nonumber
\\
\mbox{where} && D_{08} =
( G_{00} \Pi_{00} - G_{88} \Pi_{88} )^2 + 4 G_{08} \Pi_{08} ( G_{00} \Pi_{00} 
+ G_{88} \Pi_{88} ) + 4 ( G_{08}^2 \Pi_{00} \Pi_{88}
 + G_{00} G_{88}  \Pi_{08}^2).
\end{eqnarray}
In these equations $\Pi_{08}  = \frac{\sqrt{2}}{3}
(\Pi_{uu} + \Pi_{dd} - 2 \Pi_{ss})$.
For zero mixing interaction, $G_{08}=0$,
these solutions reduce to the previous ones 
for the modes $S_0$ and $S_8$, that will be denoted by 
$M_{00}$ and $M_{88}$ respectively.
The solutions for the equations (\ref{BSE-mix08}) will be labeled
$\lambda_1, \lambda_2$ or $L_1, L_2$.

\subsection{ $\sigma-f_0-a_0^0$ mixings}
\label{sec:mixing}

Next, by requiring  $f_0(980)- \sigma(500)$  mass  difference 
to be reproduced by a rotation between states $S_0$ and $S_8$
the mixing angle $\theta_{08}$ will be  calculated.
For that, a slightly different way to introduce the mesons fields will be adopted
 by  inserting the following
functional delta functions  
in  the generating functional of the model  \cite{alkofer-etal,osipov-etal}:
\begin{eqnarray} \label{deltafuncP}
1 \;  = \; \int   D [ S_i ] \; \delta \left( S_i - G_{ii}  j_{s}^i \right),
\end{eqnarray}
where $j_{s}^i = \bar{\psi} \lambda^i   \psi$,
 ($i=0,8$) provides the needed components to describe
the 
mesons and eventual mixings.
This method
 neglects possible non-factorization, which are nevertheless
  expected to be quite  small \cite{PRD-2015}.

 Consider
 the resulting (effective)
Lagrangian  terms for neutral  flavor eigenstates
 modes that 
undergo mixings,  $S_0, S_3, S_8$, with 
masses $M_{ii}^2$ for $i=0,3,8$, obtained from the BSE for 
each of the uncoupled modes.
It can be written as:
\begin{eqnarray} \label{P0P8-mix}
{\cal L}_{mix} &=&
- \frac{M_{88}^2}{2} S_8^2  
- \frac{M_{00}^2}{2} S_0^2
- \frac{M_{33}^2}{2} S_3^2
 + 
 \bar{G}_{08}    S_0 S_8 
+   \bar{G}_{03}    S_0 S_3
+   \bar{G}_{38}  S_3 S_8 .
\end{eqnarray}
The mixing terms were obtained with the introduction of the auxiliary field method,
Eq. (\ref{deltafuncP}), and the resulting coupling constants can be written as:
\begin{eqnarray} \label{mix-08}
\bar{G}_{i\neq j} = \frac{ G_{ij} }{  G_{ii} G_{jj} },
\end{eqnarray}
 For the 
leading  mixing $f_0-\sigma$,
according to the 
convention from   \cite{klempt,oller-mixing,brigitte-etal},  the rotation
 can be written as:
\begin{eqnarray} \label{mix-rot}
| f_0  > &=& 
\cos \theta_{s} | S_8 > - \sin \theta_{s} | S_0 >,
\nonumber
\\
| \sigma > &=& 
\sin \theta_{s} | S_8 > + \cos \theta_{s} | S_0 >.
\end{eqnarray}
 Note however that other works, for example
 \cite{teshima-etal},  consider a different rotation matrix.
 By performing the rotation for this mass eigenstates basis,
  and comparing  to the above $0-8$ mixing,
the following $\sigma-f_0$ mixing angle is obtained:
\begin{eqnarray} \label{angle08}
  \theta_{s}  = \frac{1}{2} \arcsin \left( 
\frac{  2 \tilde{G}_{0 8}  }{   (M_{f_0}^2 - M_{\sigma}^2 )  }  \right),
\;\;\;\;\;
\tilde{G}_{08} = \frac{\bar{G}_{08}}{G_{00} G_{88}
 \left(1 - \frac{\bar{G}_{08}^2}{G_{00}G_{88}} \right)}.
\end{eqnarray}

Next,  estimations for all the mixings $0-3-8$ ($\sigma-a_0-f_0$)
will be provided.
For the complete rotation matrix of the neutral flavor states,
 $S_3,S_8,S_0$,
to the mass eigenstates $a_0^0,\sigma, f_0$ we will adopt an analogous matrix to 
the one for the pseudoscalar mesons ($\pi_0, \eta, \eta'$) 
\cite{feldman-etal-kroll,pi0-eta-etap,JPG-2022}
because the strengths of each of the mixing follow
the same hierarchy in both channels, scalar and pseudoscalar.
The rotation matrix can be 
 decomposed into two parts: one that drives $S_8$ and $S_0$ into 
strange and non-strange basis ($S_s$ and $S_{ns}$)
 and the other
that mixes these states to form the physical mass eigenstates.
The mixing matrix $M$
can  be written as:
\[
 \left( \begin{array}{c  } \label{matrix}
a^0
\\
f_0
\\
\sigma
   \end{array} \right)
=  M   \left( \begin{array}{c  }
S_3
\\
S_8  
\\
S_0
   \end{array} \right) 
=  
 \left( \begin{array}{c c c }
1 &   \sqrt{\frac{2}{3} } 
(\varepsilon_1^s + \varepsilon_2^s C_\psi) - \frac{\varepsilon_2^s S_\psi}{\sqrt{3} }
 & \sqrt{\frac{1}{3}}   (\varepsilon_1^s + \varepsilon_2^s C_\psi) + \sqrt{\frac{2}{3}} \epsilon_2^s S_\psi
\\
-\varepsilon_2^s-\varepsilon_1^s C_\psi &  \sqrt{\frac{2}{3}} C_\psi
- \frac{S_{\psi}}{\sqrt{3}}
 &    \sqrt{\frac{2}{3}} S_\psi
+ \frac{C_{\psi}}{\sqrt{3}}
\\
-\varepsilon_1^s S_\psi &
 \sqrt{\frac{2}{3}}  S_\psi + \frac{C_\psi}{\sqrt{3}} & 
- \sqrt{\frac{2}{3}}  C_\psi + \frac{S_\psi}{\sqrt{3}}
   \end{array} \right)
 \left( \begin{array}{c  }
S_3
\\
S_8  
\\
S_0
   \end{array} \right).
\]

where $C_\psi = \cos (\psi)$ and $S_\psi = \sin (\psi)$,
being that  $\psi = \theta_{s} + arctan (\sqrt{2})$  
 - similarly to 
 \cite{delbourgo-scadron} - 
and two small mixing parameters $\varepsilon_1^s$ and $\varepsilon_2^s$.
By neglecting them ($\varepsilon_1^s = \varepsilon_2^s = 0$)
the 2 dim rotation (\ref{mix-rot}) is recovered.
Since the mixing $G_{08}$ is by far the largest one,
$\theta_s$ can be calculated from Eq. (\ref{angle08}) and used in the matrix above 
that can be used when relating both representations to yield 
analytically:
\begin{eqnarray} \label{eps12}
\varepsilon_2^s &=& 
\frac{ 
 \bar{G}_{03} 
- M_{f_0}^2 \frac{ S_\psi C_\psi}{\alpha} \bar{G}_{38}
}{ \gamma  + M_{f_0}^2 S_\psi C_\psi \frac{\beta}{\alpha}},
\nonumber
\\
\varepsilon_1^s &=&
\frac{ \bar{G}_{38} + \frac{\beta}{\gamma} \bar{G}_{03}
}{ \alpha  + M_{f_0}^2 S_\psi C_\psi \frac{\beta}{\gamma} } ,
\\
\mbox{where} && \alpha 
=
M_{f_0}^2 S_\psi^2 + M_{\sigma}^2 C_\psi^2 - M_{a_0^0}^2 ,
\nonumber
\\
&& \beta =
(M_{a_0}^2  - M_\sigma^2 ) C_\psi ,
\nonumber
\\
&& 
\gamma = - M_\sigma^2 S_\psi + M_{a_0}^2 .
\nonumber
\end{eqnarray}

\subsection{  Mixed content of   $\sigma(500)-f_0(980)-A_0^0(980)$ 
}

Next, 
it is possible to estimate the  relative 
 strengths of the transitions  $(A_0\to f_0)/(f_0\to A_0)$ 
from the mixing structure above.
  For that, firstly  
consider the  relatively larger  components 
$S_0,S_8$ and $S_3$ respectively of each of the mesons $\sigma(500), f_0(980)$ and
$A_0^0(980)$.
If the $f_0$ meson is rather a state $S_8$ then 
the mixing to $A_0^0$ (that is assumed to be mainly  a $S_3$ state)
will be ruled by the mixing coupling constant $G_{38}$ and so on.
So that one could  associate, in the leading order, this transition
to $f_0 \to S_3 \simeq A_0$.
The opposite transition should then be given, in the leading order,
by $A_0 \to S_8 \simeq f_0$.
The 
conservation of energy is important in  mixing 
processes involving   $\sigma(500)\to f_0(980)$ and $\sigma(500)\to A_0^0(980)$
that 
 would require mesons to carry non-zero momentum/kinetic energy.
  The width of the $\sigma$, nevertheless, is very large - i.e. 
$\Gamma_\sigma \sim
300$MeV -  and this may reduce 
this inconvenience.
However, there is a high expectation that $\sigma$ has strong mixing(s) with other 
types of components and
 its oscillations to $A_0^0$/$f_0$ will not be estimated.
Considering the normalized relative contributions,
the   amplitude of each of these components $S_0, S_3, S_8$ 
for each of the scalar mesons $A_0^0, f_0, \sigma$ can be written as:
\begin{eqnarray}
|A_0^0 > &=& A_{3,a_0} | S_3 > 
+ A_{8,a_0} | S_8 > + A_{0, a_0} | S_0 >,
\\
| f_0 > &=& A_{3,f_0} | S_3 > 
+ A_{8,f_0} | S_8 > + A_{0, f_0} | S_0 >,
\\
| \sigma  > &=& A_{3,\sigma} | S_3 > 
+ A_{8,\sigma} | S_8 > + A_{0, \sigma} | S_0 >.
\end{eqnarray}
For the rotation matrix given above, after normalization, these coefficients
 can be written as:
\begin{eqnarray} \label{A0-380}
A_{3,a_0} \equiv \frac{a_{0,(3)} }{a_0}  =
\frac{1 }{ \sqrt{ 1 + (\varepsilon_1^s + \varepsilon_2^s C_\psi)^2 +
(\varepsilon_2^s S_\psi)^2}}, \;\;\;\;
A_{8,a_0} \equiv
\frac{a_{0,(8)} }{a_0 } =
[ \varepsilon_1^s + \varepsilon_2^s C_\psi]
\frac{a_{0,(3)} }{a_0},
 \;\;\;\;
A_{0,a_0} \equiv
\frac{a_{0,(0)} }{a_0 } = 
 [-  \varepsilon_2^s S_\psi ]
\frac{a_{0,(3)} }{a_0} ,
\nonumber
\\
A_{3,f_0} \equiv \frac{f_{0,(3)} }{f_0} =
\frac{- \varepsilon_2^s - \varepsilon_1^s C_\psi }{ 
\sqrt{ 1  
+ (\varepsilon_2^s + \varepsilon_1^s C_\psi)^2  } }, \;\;\;\;
A_{8,f_0} \equiv \frac{f_{0,(8)} }{f_0 } =
\frac{  C_\psi  }{
\sqrt{  1 +  
 (\varepsilon_2^s + \varepsilon_1^s C_\psi)^2  }
},
\;\;\;\;
A_{0,f_0} \equiv
\frac{f_{0,(0)} }{f_0 } =
 \frac{  - S_\psi }{ 
\sqrt{ 1 + 
 (\varepsilon_2^s + \varepsilon_1^s C_\psi)^2  }
},
\nonumber
\\
A_{3,\sigma} \equiv \frac{\sigma_{(3)} }{\sigma} =
\frac{ - \varepsilon_1^s S_\psi  }{
 \sqrt{ 1 + 
(\varepsilon_1^s S_\psi)^2} },
\;\;\;\;\;\;\;\;\;\;
A_{8,\sigma}  \equiv
\frac{\sigma_{(8)} }{\sigma } =
\frac{ S_\psi }{ \sqrt{ 1 
+
(\varepsilon_1^s S_\psi)^2} } ,
\;\;\;\;\;\;\;\;\;\;\;\;\;\;\;\;
A_{0,\sigma}  \equiv
\frac{\sigma_{(0)} }{\sigma } =
 \frac{ C_\psi }{ \sqrt{ 1 
+
(\varepsilon_1^s S_\psi)^2} }.
\end{eqnarray}
The non-leading relative components of $A_0^0$ are those from  $S_0$ and $S_8$
and they are 
given by $A_{0,a_0}$ and $A_{8,a_0}$ respectively.
They can be associated to the leading  contributions for the 
probability of $A_0^0$ to oscillate into
a $\sigma$ and to a $f_0$, or rather into their quark-antiquark components.
Similarly, for the other mixings.
The width of mesons $A_0$ and $f_0(980)$ are small and
are somewhat expected to have similar values, i.e. of the order 
of $\Gamma_{a_0} \simeq 20-70$MeV, $\Gamma_{f_0} \simeq 20-35 MeV$ \cite{PDG}.
Being their difference probably small, they were neglected in this first calculation.
It has been pointed out in \cite{decaywidth-large} that only very  different 
 decay widths  lead to sizeable effects in mesons mixing.
Specifically for the $A_0-f_0$ mixings
the  leading relative strengths of the $(A_0\to f_0)/(f_0\to A_0)$ transitions 
 will be given by:
\begin{eqnarray} \label{a0f0}
R_{a0f0} \equiv \frac{A_0^0 \to S_8 \to  f_0 }{ f_0 \to S_3 \to  A_0^0 }
\sim \frac{ \left| \frac{a_{0,(8)}}{a_0} \right|^2}{
\left| \frac{ f_{0, (3)}}{f_0} \right|^2} 
\equiv 
\left| \frac{ A_{8,a_0} }{A_{3,f_0}}  \right|^2.
\end{eqnarray}
Numerical values for this ratio will be provided below.

\subsubsection{ Dynamical estimation of $f_0-A_0$ mixing ratio}

In this section, a calculation of  the $f_0-A_0$ mixing
for the dynamical  states is presented. 
These states can be written as:
\begin{eqnarray}
|a_0 (t) > = | a_0 > e^{ - i ( m_{a_0} - \frac{i \Gamma_{a_0}}{2} ) t },
\;\;\;\;\;
|f_0 (t) > = | f_0 > e^{ - i ( m_{f_0} - \frac{i \Gamma_{f_0}}{2} ) t },
\end{eqnarray}
where $m_{a_0} \simeq m_{f_0} \simeq 980$MeV are the mesons masses 
and $\Gamma_{a_0} = 20-70$ MeV and  $\Gamma_{f_0} = 20-35$MeV their widths
\cite{PDG}.
The resulting time dependent probabilities of the
initial state $|a_0>$ (or $|f_0>$) to oscillate  to
the state $|f_0>$ (or $|a_0>$), by considering the matrix 
$M$ given above and its inverse,
are given by:
\begin{eqnarray}
(f_0 \to a_0) (t) &=&  | < a_0 | f_0 (t) > |^2
= e^{-\Gamma_{f_0} t} \left(
x_{f1}  + x_{f2}  \right)^2,
\nonumber
\\
(a_0 \to f_0 )(t) &=& | < f_0 | a_0 (t) > |^2
= e^{-\Gamma_{a_0} t} \left(
x_{a1} + x_{a2}  \right)^2,
\end{eqnarray}
 where the following parameters have been defined:
\begin{eqnarray}
x_{f1} &=& \frac{ (\epsilon_2 + \epsilon_1 \cos (\Psi) ) }{ 
1 +  \epsilon_1^2  \cos (2 \Psi) + \epsilon_2^2 + 2 \epsilon_1 \epsilon_2  \cos (\Psi) } ,
\\
x_{f2} &=& 
\frac{\epsilon_1 \cos (\Psi) (\sin^2  (\Psi) - \cos^2 (\Psi) )
}{ 
1 +  \epsilon_1^2  \cos (2 \Psi) + \epsilon_2^2 + 2 \epsilon_1 \epsilon_2  \cos (\Psi) } ,
\nonumber
\\
x_{a1} &=& \frac{  \epsilon_2 \sin^2 (\Psi) + \cos (\Psi) (\epsilon_1+\epsilon_2 \cos(\Psi) ) }{ 
1 +  \epsilon_1^2  \cos (2 \Psi) + \epsilon_2^2 + 2 \epsilon_1 \epsilon_2  \cos (\Psi) } ,
\\
x_{a2} &=& 
\frac{(\epsilon_1 + \epsilon_2 \cos (\Psi) )  ( \cos(\Psi) + \epsilon_1 \epsilon_2 \sin^2 (\Psi) )}{ 
1 +  \epsilon_1^2  \cos (2 \Psi) + \epsilon_2^2 + 2 \epsilon_1 \epsilon_2  \cos (\Psi) } ,
\end{eqnarray}
The ratio between the mixing probabilities becomes:
\begin{eqnarray} \label{rdynt}
R_{dyn} (t) = e^{ - (\Gamma_{a_0} - \Gamma_{f_0} ) t }
\frac{ (x_{a1} + x_{a2} )^2 }{ (x_{f1} + x_{f2} )^2 }.
\end{eqnarray}
This ratio will be compared to results from the
Eq. (\ref{a0f0}).
Note however that the difference
between the values of widths of the two mesons may be very small 
$\Gamma_{a_0} \simeq 20-70$MeV, $\Gamma_{f_0} \simeq 20-35 MeV$ \cite{PDG}.

.

\section{ Numerical results}
\label{sec:numerics}

The set of parameters of the model  used to obtain 
numerical estimations   are given in 
Table (\ref{table:masses}).
These parameters
 were chosen  in \cite{JPG-2022}  by requiring the model
 to reproduced neutral pion and kaon masses (fitting observables). 
The corresponding coupling constants $G_{ij}$ were found
by using the  gluon effective propagators given below.
The reliability of these sets of parameters was tested by calculating several observables.
The resulting   reduced chi-square had also been calculated   and,
 for these choices $S2,S5,S6$, they are also 
presented in this Table.
$\chi_{red}^2$ were calculated for ten observables of the light pseudoscalar mesons
 after the choice
of the three quark current masses as fitting parameters.
The resulting pseudoscalar mesons masses are also presented with the corresponding 
uncertainties for the numerical calculation of the present work.
The first effective  gluon propagator (EGP)
  is  a transversal one
 extracted from 
Schwinger Dyson equations calculations  
\cite{gluon-prop-sde,SD-rainbow}.
 It  can be written as:
\begin{eqnarray} \label{gluon-prop-sde}
(S2) : \;\;\;\;\;  D_{2} (k)  = g^2 R_T (k)  =
\frac{8  \pi^2}{\omega^4} De^{-k^2/\omega^2}
+ \frac{8 \pi^2 \gamma_m E(k^2)}{ \ln
 \left[ \tau + ( 1 + k^2/\Lambda^2_{QCD} )^2 
\right]}
,
\end{eqnarray}
where
$\gamma_m=12/(33-2N_f)$, $N_f=4$, $\Lambda_{QCD}=0.234$ GeV,
$\tau=e^2-1$, $E(k^2)=[ 1- exp(-k^2/[4m_t^2])/k^2$, $m_t=0.5 GeV$,
$D= 0.55^3/\omega$ (GeV$^2$) and 
$\omega = 0.5$ GeV.
The second EGP 
 is based in an  effective  longitudinal  confining parameterization
\cite{cornwall} that will be considered for two different values of the 
gluon effective mass.
It can be  written  as:
\begin{eqnarray} \label{cornwall}
(S\alpha=S5, S6  ) : \;\;\;\;\;\;\;\;\; D_{\alpha=5,6} (k)  = g^2 R_{L,\alpha} (k)  = 
\frac{ K_F }{(k^2+ M_\alpha^2)^2} ,
\end{eqnarray} 
where 
$K_F = (0.5 \sqrt{2} \pi)^2/0.6$ GeV$^2$, as considered in previous works
\cite{PRD-2019} to describe several mesons-constituent quark 
effective coupling constants and form factors.
Different  effective gluon masses can be tested \cite{oliveira-etal,costa-etal}
such as a constant one: 
($ M_5 = 0.8$ GeV) or  a running effective mass given by:
$M_6  = M_6 (k^2)=  \frac{0.5}{1 +  k^2/\omega_6^2 }$ GeV
for $\omega_6=1$ GeV.

\begin{table}[ht]
\caption{
\small 
Set of parameters:
  Lagrangian  quark masses and three-dimensional ultraviolet  cutoff
for $G_0=10$GeV$^{-2}$ and  corresponding reduced chi-square
 taken from Ref. \cite{JPG-2022}.
These parameters were used for all the 
gluon propagators $S_{\alpha}$, $\alpha=2,5,6$.
 The fitting observables $M_{\pi^0}, M_{K^0}$ and the masses of the  corresponding
charged mesons, and the strange quark effective mass at the physical point,
 are also given for each of the effective gluon propagator adopted.
} 
\centering  
\begin{tabular}{| c | c c c c  | c | c  |}
\hline\hline  
set  of  & 
$m_u$ & $m_d$ & $m_s$ &
$\Lambda$ &  - & -
\\
parameters &  MeV  &       MeV     &    MeV  & MeV   &  - & - 
 \\
\hline 
 $S_{\alpha}$  &
3 &  7 &  133 & 680    & & 
\\
\hline \hline \hline
 & $M_{\pi^0}$  & $M_{\pi^{\pm}}$ & $M_{K^0}$ & $M_{K^\pm}$  &  $M_s^*$ & $\chi_{red}^2$
\\
\hline
$S2$ & 133.5(5) & 133.7(5) & 498.5(5) & 490(1)  &  555(1) &  103
\\
$S5$ &  134.2(5) & 134.4(5) & 498.5(5) & 491(1)    &   567(1)   &  110
\\
$S6$ &    133.7(5) & 133.9(5)  & 498.5(5) & 491(1) &   558(1)   & 103
\\[1ex] 
\hline 
\end{tabular}
\label{table:masses} 
\end{table}
\FloatBarrier

The normalized diagonal coupling constants $G_{ii}$ obtained at the physical point,
for $S2,S5,S6$ used in this work, are presented in Table (\ref{table:Gij}).
The  coupling constant, $G_{11}$, was also
included in the table, and it shows the quite different normalization of the effective
 gluon propagators (and corresponding running coupling constant that is encoded).
The uncertainties  for each set of parameters were also included, being that
$G_{11}^n$ was taken as the reference value $G_0$.

\begin{table}[ht]
\caption{
\small Numerical results for the normalized $G_{ii}^n$ (indicated simply as $G_{ii}$ in the text)
 for 
the set of parameter $S_\alpha$ and different gluon propagators ($\alpha$).
} 
\centering  
\begin{tabular}{| c | c c c c c c c | }  
\hline\hline  
SET & $G_{11}^n$  & $G_{33}^n$ & $G_{44}^n$ & 
$G_{66}^n$ & $G_{88}^n$ & $G_{00}^n$  & $G_{11}$ 
\\
& GeV$^{-2}$  &  GeV$^{-2}$ &  GeV$^{-2}$  &  GeV$^{-2}$ & GeV$^{-2}$
& GeV$^{-2}$  & GeV$^{-2}$  
\\
\hline
\hline
$S2$  & 10.00 &  10.00(5) & 9.88(5)  & 9.84(5) & 8.82(5)  & 9.30(5) & 29.7  (5)
\\
\hline
$S5$ & 10.00 &  10.00(5) & 9.92(5) & 9.89(5)  & 9.12(5) &  9.48(5) &  0.18 (5) 
\\
\hline
$S6$ &  10.00 & 10.00(5) & 9.89(5)  &  9.85(5) & 8.89(5) &  9.33(5) & 1.20 (5)
\\
\hline \hline
\end{tabular}
\label{table:Gij} 
\end{table}
\FloatBarrier

Below, 
whereas   $M_u^*$ and  $M_d^*$
are calculated always self consistently in their gap equations with flavor
dependent coupling constants $G_{ff}$, the strange quark effective mass
$M_s^*$ was considered in two situations: firstly also calculated self consistently
(s.c.)
by varying the current strange quark mass $m_{s}$ and secondly it was
    arbitrarily varied without the  self-consistency of the gap equation (n.s.c.).
Therefore, the values of $M_s^*$ in the s.c. calculations have a minimum value 
due to the DChSB.
The mixing interactions, whenever used in the calculations,
 were normalized as $G_{i\neq j}^n \equiv \frac{ G_{i j}}{G_{11}} G_0$
and will be denoted below simply as $G_{ij}$.
All the coupling constants,  in spite of  being normalized,
 will be  denoted simply as $G_{ij}$ to avoid a heavy notation.
It is important to emphasize that 
the coupling constants $G_{i=j}$ have been normalized to reproduce the
value of reference $G_0 = 10$ GeV$^{-2}$ at the flavor
symmetric point when $m_u=m_d=m_s$.
As a consequence, it is natural to assess the effect of the flavor coupling constants
as done in \cite{JPG-2022}.

\begin{figure}[ht!]
\centering
\includegraphics[width=120mm]{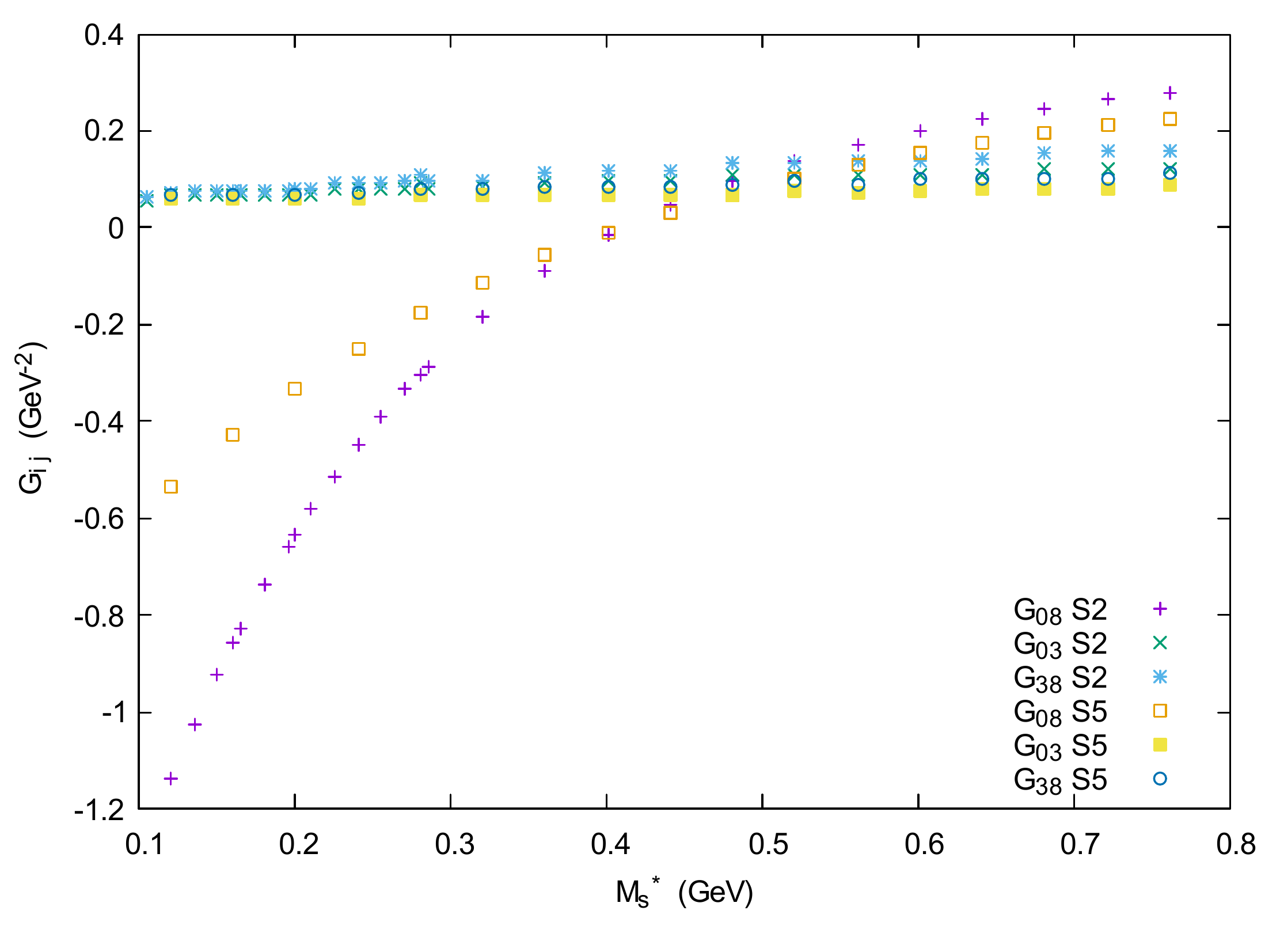}
 \caption{ 
\small
The normalized mixing coupling constants  
$G_{08} (G_0/G_{11})$, $G_{03} (G_0/G_{11})$
and $G_{38} (G_0/G_{11})$  
as functions of the strange quark mass,
for 
freely varied
$M_s^*$ and  for  two of the EGP.
}
\label{fig:G08}
\end{figure}

 In Fig. (\ref{fig:G08}) the normalized  mixing 
coupling  constants 
$G_{08}$, $G_{03}$
and $G_{38}$  
 are presented for 
 freely varying  $M_s^*$
for two of the EGP ($S2$ and $S5$).
These curves  do not depend on the current quark masses,
since the up and down quark masses are fixed.
The coupling constant $G_{08} \propto (M_s^* - M_f^*)$ for $f=u,d$
is clearly  larger than the others, that do not go to zero 
because 
 the masses $M_u^*, M_d^*$ do not assume their values at the
 symmetrical point.
This  behavior of $G_{08}$, with much larger  values,
is compatible with expectations that 
the mixing between the states $S_0,S_8$  - or correspondingly
$\sigma(500)-f_0(980)$ -  should  be larger.
Results from EGP $S6$ are  almost coincident with results from $S2$ and 
therefore
they were not shown.
An asymmetric behavior for values of $G_{ij}$ by varying
$M_s^*$ can be noticed above and below the flavor symmetric point.

\FloatBarrier

\begin{figure}[ht!]
\centering
\includegraphics[width=120mm]{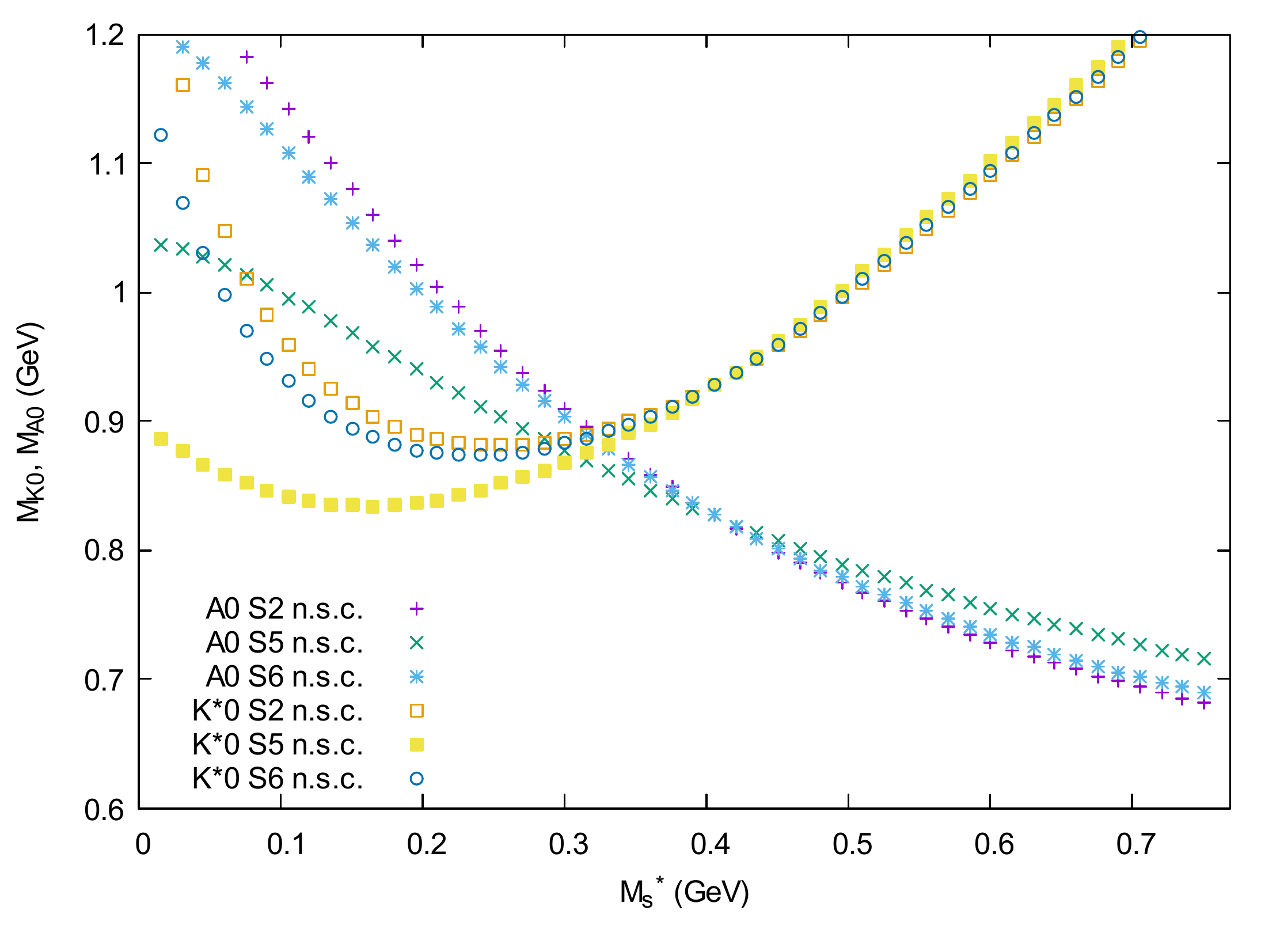}
 \caption{ 
\small
Masses of neutral $A_0$ and $K^*_0$ as functions of the strange quark mass that is 
freely varied (n.s.c.) for normalized  coupling constants $G_{ij}$ generated by
 three EGP.
}
\label{fig:MAK0}
\end{figure}

In Fig. (\ref{fig:MAK0}) the masses of 
$K_0^*(700)$ and $A_0^0(980)$ as solutions of the BSE (\ref{BSE-Gff-ii})
 are exhibited as functions of the $M_s^*$
that is varied freely  for the three EGP.
The self-consistent calculation,  for $M_s^* \geq 405$ MeV, 
yields results and behavior
 similar to the ones presented in this figure
so they were not presented.
The   $A_0$ mass reaches  the experimental value (980 MeV)
for low $M_s^*$. The mass of the $K_0^*$ improves considerably in this
range of values of $M_s^*$.
This suggests that this variation of $M_s^*$ can help to correct the mass
hierarchy problem of $K_0^*(700)-A_0(980)$ mesons.
Note however that $A_0$ does not have a strange valence quark (antiquark).
The presence of the strange valence quark (antiquark) in the $K_0^*$ 
- and correspondingly a different role for the  strange quark mass in its
 BSE - 
might contribute to the larger values of $M_{K_0^*}$
 when comparing to the experimental values \cite{PDG}.
This will be discussed further with results shown in  Table (\ref{table:masses}).
The mass difference between neutral and charged $A_0$ 
is very small, and it remains nearly the same
for the range of values of  $M_s^*$ shown in the Figure
for both the s.c. and  n.s.c. calculations.
The mass difference between neutral and charged $K_0^*$ 
depends somewhat on the EGP used to derive the coupling constants $G_{ij}$, 
although the variation with $M_s^*$ is also very small,
for both the s.c. and the n.s.c. calculations.
Therefore, the mass differences between the neutral and charged
$A_0$ and $K^*_0$ 
mesons are  not exhibited.

\begin{figure}[ht!]
\centering
\includegraphics[width=120mm]{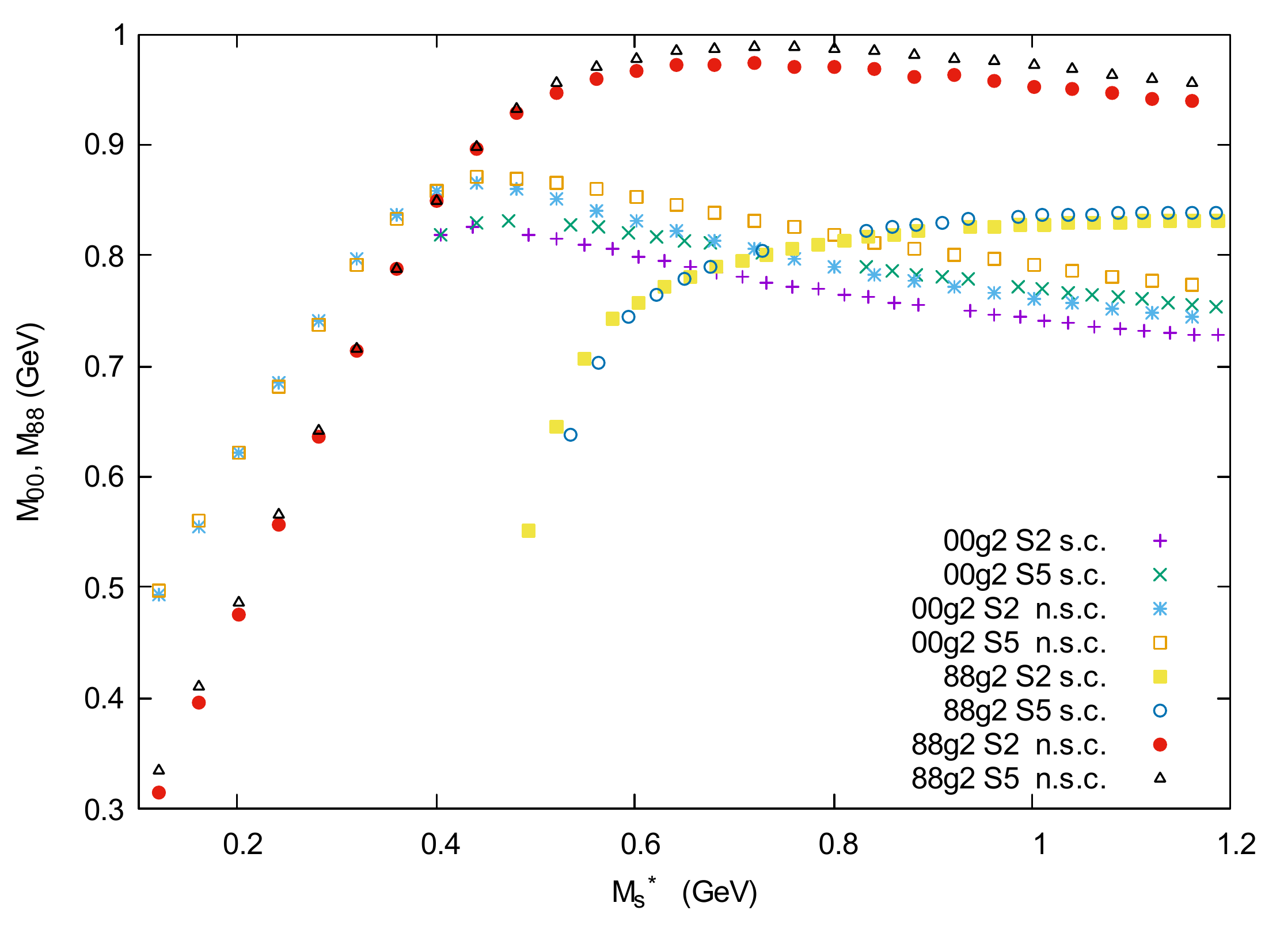}
 \caption{ 
\small
Masses of diagonal modes $S_0$ and $S_8$ as functions of the strange quark mass
that is 
freely varied for the two effective gluon propagators  - $S2$ and $S5$.
Results from both self-consistent (s.c.) and non-self-consistent (n.s.c.) calculations
are presented in the complete absence of mixings.
}
\label{fig:M00-88}
\end{figure}

The masses of the (diagonal)  modes 
$S_0$ and $S_8$,
 respectively $M_{00}$ and $M_{88}$,
are shown in Figure (\ref{fig:M00-88}) for 
two representative EGP - $S2$ and $S5$,
 for $M_s^*$ obtained in  both  ways 
(s.c.) and (n.s.c.).
The masses obtained with the set of parameters $S6$ are nearly the same as the ones 
obtained with $S2$ and  therefore were not presented.
 It is 
seen that  the variation  of $M_s^*$,
and the use of $G_{ij}$,  provide strong contributions for the 
 non degeneracy of $M_{00}, M_{88}$
that can be interpreted as  preliminary estimates for the 
masses  of the 
quark-antiquark components of the $\sigma(500)$ and $f_0(980)$
without considering   mixing states or mesons mixing that are discussed below.
Accordingly, the masses present 
the correct hierarchy  $M_{88} > M_{00}$
for the self-consistent calculation (s.c.) above a certain value of  
$M_s^{*c} \equiv M_s^* \sim  680$MeV that seems 
suitable  to describe $M_{f_0(980)} > M_{\sigma (500)}$.
Below $M_s^{*c}$ the mass hierarchy is 
the opposite, $M_{00} > M_{88}$.
For the n.s.c. calculation  the difference is quite  small, but 
the inversion of the hierarchy also 
occurs for a smaller $M_s^{*c} \sim M_0^*$,  i.e.
the flavor symmetric point.
Therefore, in this calculation, lower values of $M_s^*$ would favor 
the strange  quark-antiquark component in the $\sigma(500)$.
The choice of the effective gluon propagator does not modify considerably the results, 
being
the numerical difference  
of around $20$MeV
for large $M_s^*$. 
For small $M_s^*$, the difference due to a particular 
effective gluon propagator is much smaller.
The choice of the EGP is irrelevant around the symmetrical point
$M_u=M_d=M_s$ because of the chosen normalization of the 
coupling constants $G_{ij}$.
Overall, the EGP $S5$ leads to  larger absolute values and 
$S2$ to the smallest values.
One could not expect these curves to reach the values of the $\sigma(500)$  and 
$f_0(980)$ because of the complete  absence  of mixings to other states.
For the n.s.c. calculation both the values of the $\sigma$ and $f_0$ masses  are 
reproduced  however  at quite different 
(and unexpected) strange quark (effective) masses.

\FloatBarrier

\FloatBarrier

\begin{figure}[ht!]
\centering
\includegraphics[width=120mm]{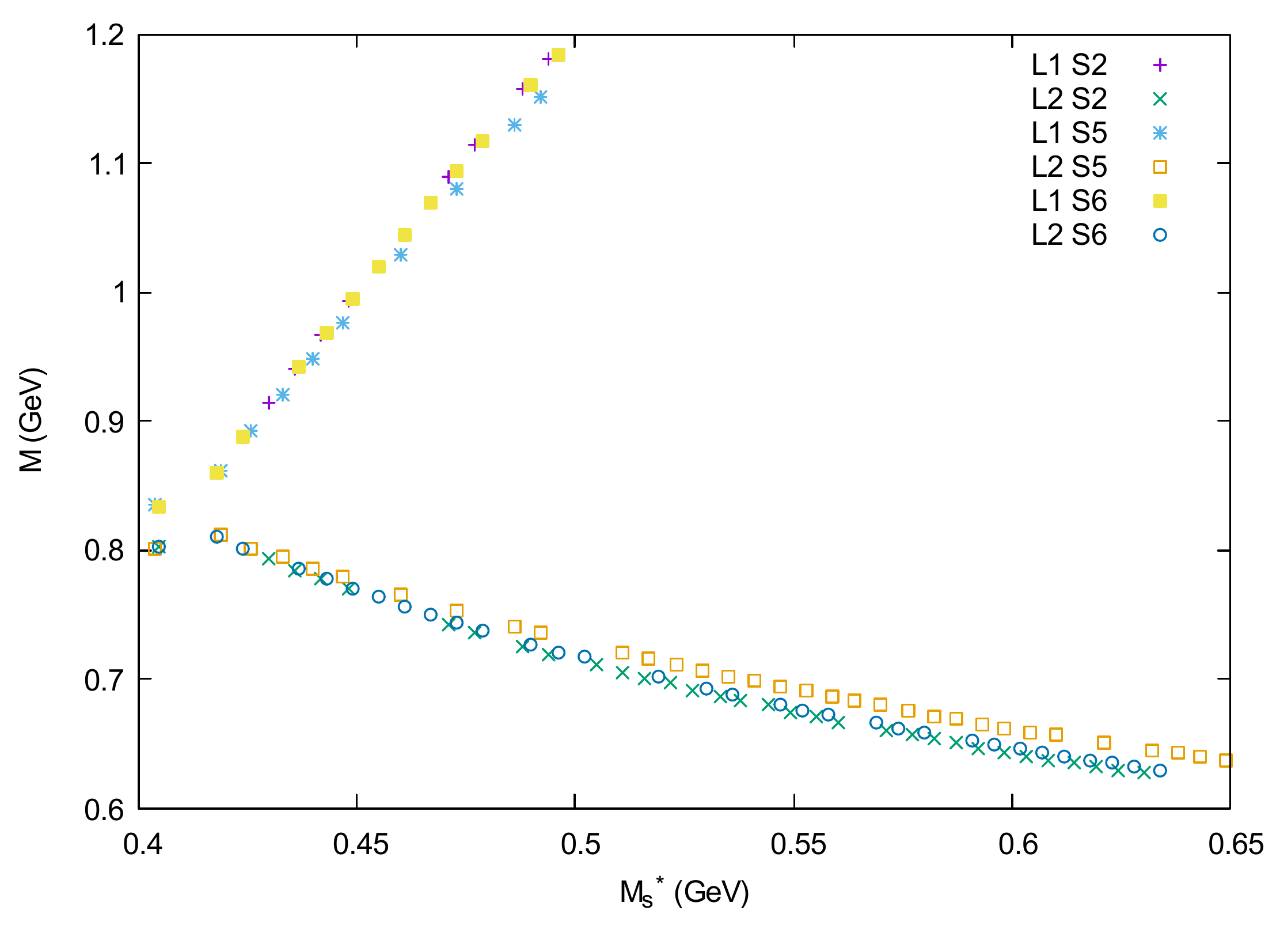}
 \caption{ 
\small
Masses of the neutral 
quark-antiquark  states, hopefully components of $\sigma(500), f_0(980)$,
obtained from Eqs. (\ref{BSE-mix08}).
They are  denoted 
respectively  as $L2, L1$ and are shown
as functions of $M_s^*$.
The three effective gluon propagators 
were considered in a s.c. calculation.
}
\label{fig:masses=25}
\end{figure}

Fig. (\ref{fig:masses=25}) presents   the 
masses of the quark-antiquark components
obtained from the coupled BSE equations (\ref{BSE-mix08}),
 with the mixing $G_{08}$,
 respectively labeled as $L2, L1$
as functions of $M_s^*$ in a
 self-consistent calculation for a range of 
current strange quark masses $0 < m_s < 150$MeV.
These states can be hopefully associated to  the components of the 
$\sigma (500)$ and  $f_0(980)$.
Coupling constants calculated with 
 the three gluon propagators $S2, S5, S6$ were 
considered. 
Results with  $S6$ are again  almost coincident with the ones for  $S2$.
The mass of the $\sigma$ is nearly  reproduced,  $M_\sigma \sim 0.6$ GeV, for
sufficiently  large $M_s^*  > 0.60$ GeV, which is   however 
in disagreement with the strange quark effective mass needed 
to provide a correct estimation for $f_0$ and 
to provide the correct mass hierarchy for 
$M_{K_0^*}, M_{a_0}$ when calculated n.s.c. as seen in Fig.   (\ref{fig:MAK0}).
If this corresponds to a relevant strangeness quark-antiquark content in the $\sigma(500)$ 
it would be somewhat "frozen"  because it would require $M_s^*$ to  be large.
However,
the $f_0(980)$ would need a quite lower $M_s^*$ to be a quark-antiquark state.

\FloatBarrier

\begin{figure}[ht!]
\centering
\includegraphics[width=120mm]{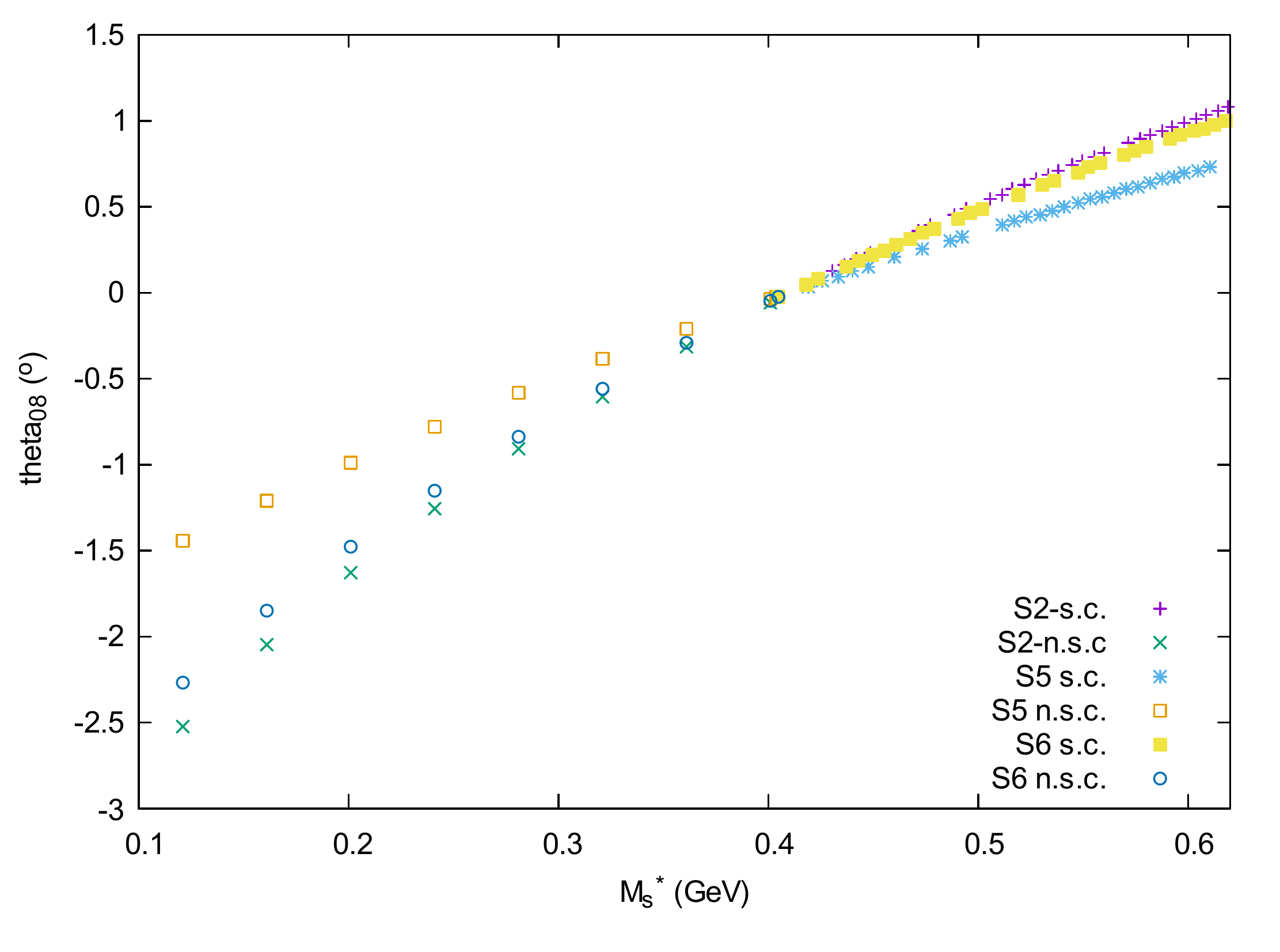}
 \caption{ 
\small
Leading mixing angle $\theta_s$ $(^\circ)$ as functions of the
 strange quark mass. 
The n.s.c. calculation  was considered for low $M_s^* < M_0^*$
and s.c. calculation  for large $M_s^* >  M_0^*$, where $M_0^*$ is the 
effective mass for degenerate quark masses.
Three effective gluon propagators were employed.
}
\label{fig:thetas}
\end{figure}

The leading mixing angle $\theta_s = \theta_{08}$,
given in Eq. (\ref{angle08}),
 that reproduces a  correct value for the  mass difference between 
$\sigma-f_0$  mesons, as a function of $M_s^*$
is presented in  Fig. (\ref{fig:thetas})
 for the cases of the three EGP,
for  s.c.  calculation
 (large $M_s^* >  M_0^*$) and for  n.s.c.
 calculation (low $M_s^* < M_0^* = 0.405$GeV)
- where $M_0^*$ is the 
effective mass for degenerate quark masses at the reference
flavor symmetric  point with  $G_0$.
The difference between s.c. and n.s.c. calculations is very small, 
 being that the s.c. only provides results 
for $M_s^* >  M_0^*$.
Besides that, the mixing angle changes sign at the symmetric point 
$M_s^*=M_d^*=M_u^*$
because $G_{08}$ does,
as seen  in Fig. (\ref{fig:G08}).
Again, the differences between results for $S2$ and $S6$ are very
small.
This mixing angle is also  the one (implicitly) responsible for the 
solutions of the coupled BSE  Eq. (\ref{BSE-mix08}) 
plotted in Fig. (\ref{fig:masses=25}).
For the sake of comparison with different mechanisms,
the mixing angle, defined in the same way,
was  $\theta_s \simeq 16^\circ-25^\circ$ (opposite sign)
in Ref. 
  \cite{brigitte-etal}  
and, for example,  $\theta_s  = (19\pm 5)^\circ$,
with some spread in the predictions obtained from different processes,
from
Ref. \cite{klempt}.

\FloatBarrier

\begin{figure}[ht!]
\centering
\includegraphics[width=120mm]{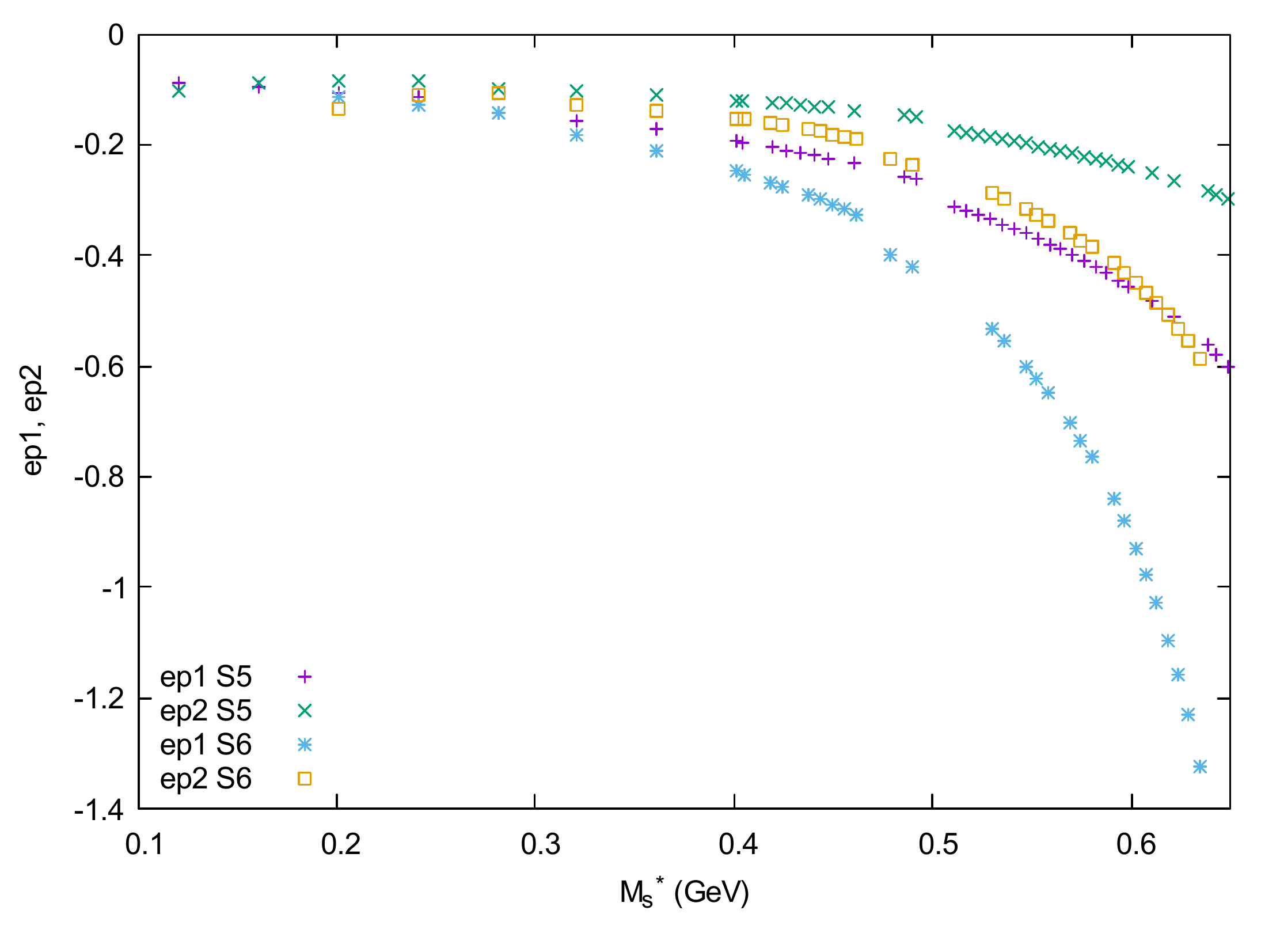}
 \caption{ 
\small
Mixing parameters $\varepsilon_1^s,\varepsilon_2^s$  as functions of the strange quark mass
for two EGP.
}
\label{fig:epsilon12}
\end{figure}

The mixing parameters $\varepsilon_1^s, \varepsilon_2^s$ 
as defined in Eq. (\ref{eps12}) are shown in Fig. 
(\ref{fig:epsilon12}) as functions of the strange quark mass.
The coupling constants  $G_{ij}$ were generated by  two of the effective gluon propagators
$S5, S6$, being that resulting points  for $S2$ are basically coincident with the ones 
from $S6$ and therefore were not plotted.
Note that these parameters are proportional to $G_{03}$ and $G_{38}$ that 
do not go to zero when $M_s^* = M_0^*$ because 
it was considered for this set of parameters that $M_u^* \neq M_d^*$.

Consider the particular hypothetical  case in which the 
quark-antiquark component of the $f_0(980)$  were
composed exclusively by a  $\bar{s}s$ state, it can  be written that:
\begin{eqnarray} \label{f0-ss}
  |f_0 > = \pm |\bar{s} s> = \pm
\frac{1}{ \sqrt{3}} \left(  | S_0 > - \sqrt{2} | S_8 > \right),
\end{eqnarray}
To obtain this state with the rotation given above by starting
from $S_8$   and $S_0$ structures
one should have:
\begin{eqnarray} \label{cos-f0-ss}
\cos (\psi) = -\frac{\varepsilon_2^s}{\varepsilon_1^s},  \;\;\;\;
\tan (\psi) =  \frac{1}{\sqrt{2}}  \to \psi \simeq  35.2 ^\circ , 
\;\;\;\; \theta_s \simeq - 19.5 ^\circ , \;\;\;\; \mbox{and} \;\;\;
\frac{\varepsilon_2^s}{\varepsilon_1^s} \simeq  - 0.82  .
\end{eqnarray}
 Results  shown in the Fig.  (\ref{fig:thetas})
are far from the limit in which this  $f_0(980)$-quark-antiquark state
 becomes exclusively
composed by $\bar{s}s$ state
because it would require,
from Eq.  (\ref{cos-f0-ss}), 
a negative ratio $\varepsilon_1^s/\varepsilon_2^s < 0$.
It can be concluded that  the  mechanism investigated in the present work, alone,
 does not lead to a  $f_0(980)$
as a $\bar{s}s$ state for the states defined in (\ref{S0S8}).

\FloatBarrier

In Table (\ref{table:masses})
some values of  masses of the 
quark-antiquark states defined above associated to the neutral mesons 
 $K_0^*(700), A_0^0(980)$
and $\sigma(500), f_0(980)$ (solutions $L1, L2$),
as defined respectively 
from Eqs.  (\ref{BSE-Gff-ii})
and (\ref{BSE-mix08}),
extracted from the figures above  are shown.
These values were chosen because they are the closest values 
to the experimental mesons masses.
 The corresponding values  
 of $M_s^*$ for these choices 
are also shown inside parenthesis.
Results for two effective gluon propagators, $S2$ and $S5$, were considered
for either  s.c. or n.s.c.
The "best  values" obtained in PDG \cite{PDG} are shown with also
masses obtained in fits of PDG  with Breit-Wigner function which are, in average, 
larger than the "best values" for these neutral mesons  except for the $f_0(980)$.
 The    uncertainties  are also included:
note that the uncertainty in the n.s.c. for the strange quark effective mass 
is  basically due to  the step in which $M_s^*$ was varied, whereas 
in the s.c. the variation of the current strange  quark  mass
lead to a considerably smaller variation of $M_s^*$.
In both cases, the uncertainty coming from the steps in variation of the 
strange quark effective masses
is  larger than the ones coming from error propagation.
For each of the s.c. and n.s.c. calculation, 
 average values for each of the meson
(and quark) masses
  are  shown in horizontal lines identified with $*$.
The low values of $M_s^*$ favors the 
correct mass hierarchy of the $K_0^*-A_0$  
and also a reasonable value for the $\sigma$ but not
the observed value for the mass of the $f_0(980)$.
For reasonably larger $M_s^*$ reasonably good values 
for the masses of quark-antiquark components
 of the $f_0(980)$
and  of the $\sigma$ are found. 
Although the corresponding  $M_s^*$ are somewhat
different, it is not clear whether another set of parameters of the 
model can improve this description,
 keeping a correct behavior of the pseudoscalar mesons masses.
A possible important  source of problem in these estimations 
should be related to the presence of the valence strange  quarks
in some mesons/states but not in others.
One can even envisage if  the coupling to other states (or mechanisms)
 can modify 
the quark gap equations - or BSE - with corrections due to gluon states, as 
for example in Ref. \cite{PRD-2014} (or \cite{PRD-2021}).

 In different versions of the  U(3)  LSM, such as
 \cite{scadron-etal,schumacher,qqlsm1,qqlsm2,tornqvist},
 results are reasonably compatible with the full 
quark-antiquark nonet structure.
In these models, the problem in the mass hierarchy usually does not appear.
 However, 
in  several versions of the LSM model, 
the mass of $a_0(980)$ may be somewhat larger than the experimental one
and the  $\kappa (K_0^*)$ may be associated to a heavier one $K_0^*(1430)$.
 In view of these results,
 it would   become somewhat
 appealing to propose  a scenario where the strange quark effective mass
 can be quite different
in each meson, due to boundary conditions or 
further dynamical couplings,  such as to provide the 
observable meson masses.
In the interval of $130 \leq M_s^* \leq  255$ MeV it is possible to fit 
the $\sigma(500), K_0^*(800)$ and $a_0(980)$ but not the $f_0(980)$ what 
would go along results from extended NJL model with higher order interactions \cite{brigitte-etal}.
In these extended models the up-down and strange quark effective masses
still may be quite different, for example 
$M_u^* \sim 244-371$MeV and $M_s^* \sim 486-646$ MeV
 \cite{brigitte-etal,hiller-etal2007},
being this difference smaller than in other (standard) versions of the U(3) NJL model.
Different results, for varying all quark masses, were obtained in \cite{santowsky-fischer}
where the large quark masses regime favor the formation of quark-antiquark meson states.
However,  we will rather emphasize that the $M_s^*$ needed for that, specially for 
the $\sigma(600)$ and $K_0^*,a_0$ states, are much lower than and different from
expected, and therefore it becomes 
somewhat artificial to claim quark-antiquark structure 
for the full nonet based exclusively  in these results.
 The most natural - although not necessary -
conclusion is that these results, extracted from the NJL model 
with flavor dependent coupling constants,  corroborate 
expectations that these 
mesons cannot be pure  quark-antiquark states.

\begin{table}[ht]
\caption{
\small  Masses of the 
 neutral quarK-antiquark mesonic states -extracted from the figures above - 
calculated at specific strange quark effective mass
- indicated in parenthesis -
for either self-consistent (s.c.) or non-self-consistent (n.s.c.),
for  calculations described in the 
text.
 $M_{00},M_{88}$ are the masses of the uncoupled  states $S_0$ and $S_8$
obtained in Eq. (\ref{BSE-Gff-ii}) and
$M_{L1,L2}$ are the coupled  mass states  by $G_{08}$ as
 defined in Eq. (\ref{BSE-mix08}).
The masses $M_{K^*_0}$ and $M_{a_0}$ were calculated with
Eq. (\ref{BSE-Gff-ii}).
Exp. are the "best fit" of the Particle Data Book Group  and 
BW $^{\dagger}$  
are  the masses from  Breit-Wigner fits   \cite{PDG}.
All the quark or meson masses obtained are shown with the corresponding uncertainty of the 
calculation.
} 
\centering  
\begin{tabular}{| c  | c  |  c |  c | c  c  |  }  
\hline  
SET - state       
   & $M_\sigma$   \; ($M_s^*$) &
 $M_{f_0}$  \; ($M_s^*$) &  
$M_{K_0^*}$ \; ($M_s^*$) & $M_{a_0}$ \; ($M_s^*$)  &
\\
 & 
 MeV \;  (MeV) &    MeV  \;  (MeV) &  MeV \; (MeV)
  &    MeV    \;  (MeV)  & 
\\
\hline
\hline
$S_{2}$ \; n.s.c.\; $M_{00}$  & 600(25) \; (175(20))  &   866(25) \; (436(20))  & - & & 
\\
$S_{5}$ \; n.s.c.\; $M_{00}$ & 600(25) \; (178(20))   & 870(25) \; (451(20)) &  -  & & 
\\
\hline
$S_{2}$ \; n.s.c.\; $M_{88}$ &  600(25) \; (185(20))  & 980(25) \; (511(20)) & - & & 
\\
$S_{5}$ \; n.s.c.\; $M_{88}$ &  600(25) \; (180(20))   &  980(25) \; (485(20)) & -    & & 
\\ 
\hline
* n.s.c. $M_{ii}$  &  600(30) \; (180(25)) & 924(60) \; (471(40)) & - & & 
\\
\hline\hline
$S_{2}$ \; s.c.\;  $M_{L1}$ &  -  & 980(25) \; (445(8)) & -   & &  
\\
$S_{5}$ \; s.c.\; $M_{L1}$ &  - &   980(25) \; (450(8)) &  -  & & 
\\
\hline
$S_{2}$ \; s.c.\;  $M_{L2}$ &  630(25) \; (630(8))  & - & - &  - & 
\\
$S_{5}$ \; s.c.\; $M_{L2}$ &  600(25) \; (649(8)) &  - &  - &  - & 
\\
\hline
* s.c. $M_{L1,L2}$ & 615(40) \; (640(18)) &  980(30) \; (448(11)) &  - & - & 
\\
\hline\hline
$S2$ \; n.s..c.   & -  & -  &      882(25) \; (255(20)) & 980(25) \; (232(20))   & 
\\
$S5$ \; n.s.c.   & -  & -  &    834(25) \; (166(20))  &   980(25) \; (130(20)) & 
\\
\hline
* n.s.c.  & - & - & 858(49) \; (211(65)) &   980(30) \; (181(70)) & 
\\
\hline\hline
Exp. \cite{PDG} &  400-550   \; (-)  
& 990 \; (-)  & 630-730 \; (-)    &  980 \; (-)   &
\\
BW $^{\dagger}$ \cite{PDG} &  400-800    \; (-)  
& 977-985    \; (-)   &   845     \; (-)   &  982-993    \; (-)   &  
\\\hline  
\end{tabular}
\label{table:BSE}  
\end{table}
\FloatBarrier

The relative composition of the neutral  mesons  that undergo mixing
according to the rotations given in Section (\ref{sec:mixing})
were defined in Eqs. 
(\ref{A0-380})
and they are shown in Table (\ref{table:composition} 
for varying values of $M_s^*$  with a  n.s.c. calculation.
It is interesting to notice 
the quark-antiquark components of the  $f_0$ and $\sigma$ 
can be said to  become different with varying $M_s^*$
with the inversion of the roles  of the states $S_0$ and $S_8$.
It is important to emphasize however that
these two meson states have valence strange quark/antiquark
according to the definitions of $S_0$ and $S_8$.
Numerical values for the ratio
Eq. (\ref{a0f0}), 
$R_{a0f0} \equiv \frac{a_0^0 \to S_8 \to  f_0 }{ f_0 \to S_3 \to a_0^0 }$,
are  also given in the next-to-last column of the Table.
Exp. shows two different estimations  - model I and model II -
for  the
ratio of the  intensity of the mixings
$(a_0 \to f_0)/(f_0\to a_0)$ 
found in BESS-III - see  Table II
of  Ref. \cite{a0-f0-mix-exp}.
These  resulting numerical values (I) and (II) are reproduced either for strange quark
effective mass around the symmetrical point or very large 
$M_s^* \sim 0.630$ GeV - calculation (I) 
or for very small $M_s^* \sim 0.24$GeV - calculation  (II).
From these estimations
 the $a_0-f_0$ mixing is compatible with  quark-antiquark  states mixing
 although other contributions from non-quark-antiquark 
for the structure of  $f_0(980)$ can contribute in the same way
by maintaining these contributions correct.
Some estimates for the $a_0-f_0$ mixing, for both 
cases quark-antiquark or tetraquark structures, 
  provide smaller strengths
 \cite{kirk-etal,close-tornq,achasov-etal}
and other reproduce one of these two results, (I)  or  (II) from BESSIII,
  \cite{oller-meiss-,1808.08843,wang-etal}.

In the last column of the Table there are results for 
the $t=0$ limit of 
Eq. (\ref{rdynt}) as  a function of $M_s^*$.
For this figure, it was adopted
 $m_{a_0}= m_{f_0} = 980$ MeV and $\Gamma_{f_0} = 30$MeV and
$\Gamma_{a_0}=70$MeV.
This dynamical
calculation presents  the opposite behavior of the 
results obtained from Eqs.  (\ref{A0-380}) for most part of the range of values of
$M_s^*$:
as the strange quark effective mass increases, the ratio $R_{dyn} (t=0)$
increases. 
However, for larger values of $M_s^*$ (above 0.593 GeV)
 the behavior  becomes similar to 
the first calculation (Eq. (\ref{A0-380})
 changes to the same of $R_{dyn}$
 with 
very different absolute values.
The reason for this discrepancy is the fact that in the 
first calculation, Eqs. (\ref{A0-380}), the mixings were considered
only as the leading contributions of intermediary states
$S_8$ or $S_3$.

\begin{figure}[ht!]
\centering
\includegraphics[width=120mm]{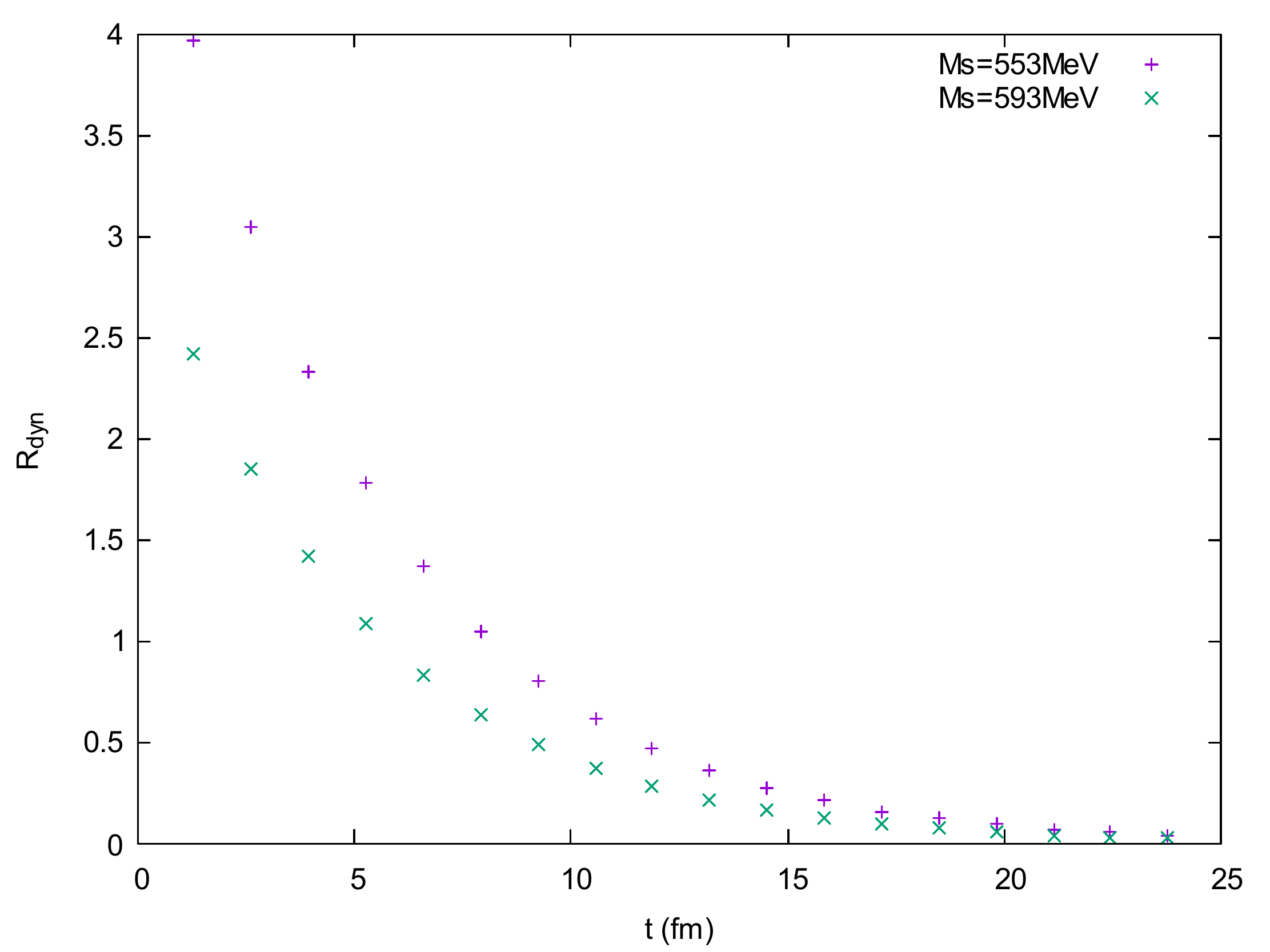}
 \caption{ 
\small
Temporal evolution of eq. (\ref{rdynt})
for the set of  parameters S5 above and two values of the strange quark mass
effective mass for $m_{a_0}=m_{f_0}=980$MeV
$\Gamma_{f_0} = 30$MeV and
$\Gamma_{a_0}=70$MeV.
}
\label{fig:rdynt}
\end{figure}

In Fig. (\ref{fig:rdynt})
the temporal evolution of the ratio of mixing probabilities 
of eq. (\ref{rdynt}) is presented for the parameters of the Set 5  shown above
by considering
$\Gamma_{f_0} = 30$MeV and
$\Gamma_{a_0}=70$MeV 
and two different values of the strange quark effective mass:
\begin{eqnarray} \label{rdynt-par}
M_s^* = 553 \; MeV \; \mbox{and} \; 593 \; MeV.
\end{eqnarray}
There is however a large uncertainty in the experimental value of the 
width of the meson $A_0(980)$ and therefore this timescale of decreasing
$R_{dyn}(t)$ may be considerably different (larger).

\FloatBarrier

\begin{table}[ht]
\caption{
\small 
Relative composition of each of the neutral quark-antiquark states
that possibly belong to the  scalar mesons
$a_0, f_0$ and $\sigma$ - as proposed in Eqs. (\ref{A0-380}) - 
as functions of the 
strange quark effective mass for  the effective 
 gluon propagator $S5$.
In the   column - $R_{a0f0}$, the ratio given by Eq. (\ref{a0f0}), 
$R_{a0f0} \equiv \frac{a_0^0 \to S_8 \to  f_0 }{ f_0 \to S_3 \to a_0^0 }$.
 In the last column, the dynamical calculation, $R_{dyn}$, at  $t=0$ for 
$\Gamma_{f_0} = 30$MeV,
$\Gamma_{a_0}=70$MeV, and $m_{a_0}= m_{f_0}=980$MeV for eq. (\ref{rdynt}).
In the last line two different estimations 
for  the relative intensity of mixings
obtained from the experimental observation in Ref. \cite{a0-f0-mix-exp}.
} 
\centering  
\begin{tabular}{ | c || c c c | c  c c | c c c | c ||| c |}  
\hline\hline  
$M_s^*$  (GeV)   & 
$A_{3,a_0}$ & $A_{8,a_0}$ & $A_{0,a_0}$ & 
$A_{3,f_0}$ & $A_{8,f_0}$ & $A_{0,f_0}$ & 
$A_{3,\sigma}$ & $A_{8,\sigma}$ & $A_{0,\sigma}$ & $R_{a0f0}$  & $R_{dyn}(t=0)$ 
 \\
\hline 
    0.201 &  0.97 &  -0.18 & -0.003 &
 0.18  & 0.96 &  0.04  &   0.004  & -0.04  & 0.99 &  0.99 & 0.080
\\
  0.241 &   0.96 &  -0.19 &   0.014 &  
 0.19  & 0.95  & -  0.16 & 
 -0.02  & 0.17  & 0.98 & 0.97 & 0.071
\\
 0.281 & 0.95 & -0.22 &  0.034 & 
 0.22  & 0.88  & -0.35 & 
-0.05  & 0.36 &  0.93 & 0.88  &  0.070
\\
0.321  & 0.94  & -0.23  & 0.05 &  
 0.22  & 0.80  & -0.51
& -0.08  & 0.53 &   0.84 & 0.75 &  0.073
\\
0.361 &   0.93 &  -0.24 &   0.07 & 
0.22  & 0.69 & -0.64 & 
-0.11 & 0.67  & 0.72  & 0.59 &  0.084
\\
0.401 &  0.93   & -0.25   & 0.09 & 
 0.22 &  0.58  & -0.75
& -0.15 &   0.77 &   0.59 & 0.43 &  0.101
\\
\hline
 0.405  & 0.92   & -0.25   & 0.09 & 
 0.22  & 0.57  & -0.76 & 
-0.15 &  0.78   & 0.58 & 0.41 &  0.103
\\
  0.433 &  0.92 &  -0.25 &   0.10 & 
 0.22  & 0.47   & -0.82 & 
-0.18  &  0.84   & 0.48  & 0.30 &  0.124
\\
  0.460 &   0.91 &  -0.26 &   0.11 & 
 0.22  & 0.38  & -0.87 & 
-0.20  & 0.87  & 0.36  & 0.20 &  0.154
\\
  0.486 &   0.90 &  -0.27 &   0.13 & 
 0.22  & 0.29 & -0.90 &  
-0.23  &  0.90  &  0.29 & 0.12 &  0.196
\\
  0.511 &   0.87 &  -0.30 &   0.15 & 
 0.23  & 0.21  & -0.92 & 
-0.28  & 0.89  & 0.20 & 0.05 &  0.248
\\
  0.529 &  0.86 &  -0.31 &  0.16 & 
 0.23  & 0.15 & -0.93 &
-0.30  & 0.89  & 0.14  & 0.02 &   0.311
\\
  0.553 &   0.84 &  -0.32 &   0.17 & 
 0.22   & 0.09 & -0.94 & 
-0.32  & 0.88  & 0.09 & 0.0 &  0.448
\\
   0.559 &   0.84 &  -0.33 &   0.17 & 
0.22  & 0.05 & -0.95 & 
-0.33   & 0.87   & 0.05 &  0.001 &  0.498
\\
  0.576 &   0.82 &  -0.34 &   0.18 & 
 0.21  & -0.002 & -0.95 & 
-0.35 &   0.86  & -0.002 & 0.02 &  0.703
\\
  0.593 &   0.80 &  -0.35 &   0.19 & 
 0.20  & -0.06 & -0.96 & 
 -0.37  & 0.83 & -0.05 & 0.08 &  1.086
\\
  0.610 &   0.79 &  -0.36 &   0.20 & 
0.19   & -0.11  & -0.96 & 
-0.39 & 0.81  & -0.09 & 0.19 &  1.886
\\
  0.632 &   0.79 &  -0.35 &   0.19 & 
 0.16  & -0.18  & -0.96 &
-0.39  & 0.80  & -0.15 & 0.46  & 7.570
 \\[1ex] 
\hline 
Exp. \cite{a0-f0-mix-exp}
 & & & & & & & & & & (I) \;  0.404  & (I) \;  0.404 
\\
 & & & & & & & & & & (II) \;  0.97 & (II) \;  0.97
\\
\hline
\end{tabular}
\label{table:composition} 
\end{table}
\FloatBarrier

\section{ Conclusions}

In this work,  the mass spectrum  of  scalar quark-antiquark states
as possible
components  of the 
 lightest scalar mesons, $\sigma(500)$, $K_0^*(700)$, $A_0(980)$
and $f_0(980)$,  were investigated by 
considering   the 
NJL model with flavor dependent coupling constants.
The parameters of the model were fitted in 
Ref. \cite{JPG-2022} to reproduce the neutral
pion and kaon masses
and for which several observables had been
investigated.
The mixing coupling constants $G_{i \neq j}$ - and $G_{f_1\neq f_2}$ -
 were not really considered for 
most of the calculations, and they will be  treated in a complete way
in a separate work.
The flavor dependent coupling constants $G_{ii}$, 
and conversely $G_{ff}$,
extracted from vacuum polarization with non-degenerate
quark masses,  still  do
not provide exactly the lightest  scalar mesons spectrum as 
quark-antiquark states.
The role of the strange quark and its mass,
both as a valence and sea quarks, was further investigated
by 
varying the strange quark 
effective mass  in two ways:
firstly within a self-consistent calculation for all the 
 gap equations 
and secondly with 
a non-self-consistent calculation.
In the first case, the resulting strange quark effective mass
is limited to values $M_s^* \geq M_0^* \simeq 405$MeV
due to DChSB.
In the second case different values were possible for 
$M_s^*  > 0$.
In both cases, in all steps, the up and down quark effective masses
were calculated self consistently from their gap equations.
The mesons $K_0^*(700)$ and $A_0(980)$ have their mass hierarchy 
modified by considering large  variations of $M_s^*$:
The mass of $A_0(980)$ ($K_0^*(700)$)
 increases (decreases) for decreasing $M_s^*$. 
They nearly 
reach  the  corresponding experimental values   of the  meson masses
 at slightly 
different lower values of the strange quark effective mass.

The mass of the $K_0^*(700)$ obtained
by varying $M_s^*$,  however,
 is only close to the Breit-Wigner mass ($845$MeV) of  \cite{PDG}.
It is important to emphasize  that $A_0(980)$ 
does not have a strange valence  quark
whereas $K_0^*(700)$ has a strange valence quark or antiquark
 and this comparison may  suggest some
difference in the  self-consistent dynamics of the sea and valence quarks.

The flavor eigenstates  $S_0,S_8$ were 
considered as responsible for the main quark-antiquark content
of the $\sigma(500)$ and $f_0(980)$
and the corresponding masses
were calculated as functions of the $M_s^*$.
Firstly, this was done  for the uncoupled BSE. 
As a result,
 the mass non-degeneracy 
$M_{00}-M_{88}$
becomes closer to  $M_\sigma-M_{f_0}$
for very large $M_s^* > 1$ GeV 
of an s.c. calculation, 
and this is 
not enough to pin down the experimental values.
The mass hierarchy of $M_{88}-M_{00}$ changes,
in both calculations s.c. and n.s.c., at different values of $M_s^*$
being $M_{00} > M_{88}$ for sufficient low values of $M_s^*$.
Although it may be missing some mixing with further
non-quark-antiquark states, this inversion
of mass  hierarchy  is interesting.
By including the mixing between the 
flavor eigenstates $S_0$ and $S_8$
by means of the coupling constant, $G_{08}$,
results improve considerably to describe $M_\sigma$ and $M_{f_0}$.
However, for the sets of parameters adopted in this work
and for the strengths of the coupling constant $G_{08}$, 
the experimental values of their masses  are found for 
different values of  $M_s^*$, i.e.
$\sigma(500)$ mass is somewhat reproduced for $M_s^* \sim 640$MeV
(when $M_\sigma \sim 620$MeV)
and $f_0(980)$ for $M_s^* \sim 450$MeV.
Specific results for the $S_0,S_8$ states
 and all the mesons 
are somewhat summarized in Table (\ref{table:BSE}).
Similarly to Ref. \cite{klempt}
we can envisage further mixing with non-quark-antiquark states.
However, there are  ambiguities in defining couplings and coupling constants
so that this further step to describe mesons masses will not be performed in the 
present work.
 It could be envisaged from  these results that 
the lightest scalar mesons, as defined in (\ref{scalars-structure}), could be
(mostly) formed by quark-antiquark states if the strange quark mass for each of these mesons
were (very) different. However, this assumption seems  somewhat artificial.

The  parameterization usually adopted for the pseudoscalar mesons
 $\eta-\eta'-\pi^0$  mixing problem,
as proposed in Ref. \cite{JPG-2022},
was adopted for the neutral mesons mixings $f_0-A_0-\sigma$.
The coupling constants $G_{08}, G_{03}$ and $G_{38}$
are mostly relevant respectively for 
$f_0-\sigma$, $A_0-\sigma$ and $A_0-f_0$ mixings, being
the leading mixing angle $\theta_{08}$  (from $G_{08}$).
This angle was calculated to reproduce the
$f_0-\sigma$ mass difference.
The mixing angle $\theta_{08}$ changes sign at the 
symmetric point because it is proportional to 
$G_{08} \propto (M_s^* - M_u^*)$.
The other  mixing  coupling constants are smaller since
they are, in the leading order,  proportional to  smaller 
quark masses  differences, $G_{03}, G_{38}  \propto (M_d^* - M_u^*)$,
 that was kept fixed.
 The meson $f_0(980)$ cannot be described as a $\bar{s}s$ state
for the interactions 
and set of parameters discussed in this work. 
The  parameters $\varepsilon_1^s, \varepsilon_2^s$
were calculated to estimate the
 $A_0^0 \to f_0$ and $f_0\to A_0^0$ mixings.
Two ways of estimating the ratio  $(A_0 \to f_0)/(f_0 \to A_0)$ were 
presented and 
  compared to the estimation from experimental results 
 obtained by BESS-III collaboration  \cite{a0-f0-mix-exp}.
 The first calculation picks up particular intermediary 
states $S_3$ and $S_8$.
The second calculation is a dynamical one,
for which the mixing should disappear at a timescale of  nearly $15-20$ fm
for $\Gamma_{f_0} = 30$MeV and
$\Gamma_{a_0}=70$MeV. 
Different (smaller)  values  for $\Gamma_{a_0}$ will lead 
to very different  (larger) time scales for the decrease of  this ratio $R_{dyn}(t)$.
The strength of the $t=0$ mixing probabilities of the two calculations 
have slightly different behaviors with respect to the variation of the strange quark effective mass.
The   different values for the ratio $R_{A0f0}$
can be obtained by the calculation above
for different values of $M_s^*$.
In conclusion, the above results  may shed some further
light on the intricate problem of the structure of the  lightest scalar mesons.
The failure in reproducing the scalar meson spectrum,
in agreement with several other approaches,
 may be further an indication of the reliability of the model.
The complete investigation of the   effects of 
the mixing interactions
and the corresponding decays of the quark-antiquark states 
that contribute for the mesons decay that were  not be addressed
in this work,
 are left for further work.

\section*{Acknowledgements}

F.L.B. thanks short conversation with C.D. Roberts
and  partial financial support from
CNPq-312750/2021-8.
The author is a member of the INCT-FNA
Proc. 464898/2014-5.

\end{document}